%% file: thesis_arXiv.tex
\documentclass[11pt,a4paper]{phdthesis}
\pdfoutput=1
\usepackage{times}

\usepackage{fancyhdr} 
\usepackage{xspace}
\usepackage{setspace}
\usepackage{enumitem}
\usepackage[toc,page]{appendix}
\usepackage[usenames, dvipsnames]{color}
\usepackage[dvipsnames]{xcolor}
\usepackage{quoting}
\usepackage{hyperref}

\usepackage{makeidx}     
\makeindex

\usepackage{amsmath}
\usepackage{amssymb}
\usepackage{amsthm}
\usepackage{epsfig}
\usepackage{graphics}
\usepackage{float}
\usepackage{capt-of}
\usepackage{here}
\usepackage{sectionapp}
\usepackage{aas_macros} 
\usepackage{comment}
\usepackage{epigraph}
\usepackage{natbib}
\bibpunct{(}{)}{;}{a}{}{,}
\usepackage{threeparttable}
\usepackage[referable]{threeparttablex}
\usepackage{tabularx}
\usepackage{array}
\newcolumntype{C}{>{\centering\arraybackslash}X}
\usepackage{dcolumn}
\newcolumntype{d}[1]{D{.}{.}{#1}}
\usepackage{longtable}
\usepackage[english]{babel}
\usepackage{microtype}
\usepackage{wallpaper}
\usepackage{subfigure}
\usepackage{wrapfig}
\usepackage{booktabs}
\usepackage{multirow}
\usepackage[outer=25mm,inner=25mm,top=25mm,bottom=25mm]{geometry}
\usepackage{upgreek}
\usepackage[strict]{changepage}
\usepackage[labelsep=colon, labelfont=bf, font=small]{caption}
\usepackage{graphicx}
\usepackage{placeins}

\definecolor{mygray}{rgb}{0.5,0.5,0.5}
\usepackage{listings}
\floatstyle{plaintop}
\restylefloat{table}

\graphicspath{{figures/}}
\input{var.tex}

\begin{document}

\include{divisions/thesisfront}

\frontmatter

\include{divisions/abstract}

\include{divisions/preface}

\include{divisions/ack}

\pagestyle{fancy}
\rfoot{\thepage}
\lhead{}
\rfoot{}
\rhead{}
\renewcommand{\headrulewidth}{0pt}

\thispagestyle{empty}
\addtocontents{toc}{\protect\thispagestyle{empty}}
\tableofcontents


\mainmatter	

\include{divisions/intro_DLA}

\include{divisions/intro_grism}


\include{divisions/Krogager2012}

\include{divisions/Krogager2013}

\include{divisions/redQSOs}

\include{divisions/HAQ}
\include{divisions/Krogager2015}

\include{divisions/Krogager2014}

\include{divisions/summary/conclusion}

\include{divisions/summary/summary_dk}

\include{divisions/summary/summary_es}

\begin{appendices}

\include{divisions/appendix/appendix1}

\include{divisions/appendix/publications}

\end{appendices}

\bibliographystyle{apj_3}

\end{document}

%% file: var.tex
\graphicspath{{figures/}}

\newcommand\ion[2]{#1\,{\sc #2}}
\newcommand{\farcs}{\mbox{\ensuremath{.\!\!^{\prime\prime}}}}
\newcommand{\lya}{Ly$\alpha$}

\def\Hcrh2{\mbox{$\Gamma_{\rm{CR,H}_2}$}\xspace}

\def\Ch2{\mbox{$\Lambda_{\rm{H}_2}$}\xspace}

\def\NHI{\mbox{$N_{\rm{H}\textsc{i}}$}\xspace}

\def\nh2{\mbox{$n_{\rm{H}_2}$}\xspace}

\def\NH2{\mbox{$N_{\rm{H}_2}$}\xspace}

\def\Mh2{\mbox{$M_{\rm{H}_2}$}\xspace}

\def\dv{\mbox{$\Delta V_{90}$}\xspace}

\def\g0{\mbox{$G_0$}\xspace}

\def\Av{\mbox{$A(V)$}\xspace}

\def\Sh2{\mbox{$\Sigma_{\rm{mol}}$}}

\def\rh2{\mbox{$R_{\rm{H}_2}$}\xspace}
\def\rh2{\mbox{$R_{\rm{H}_2}$}\xspace}
\def\rrh2{\mbox{$r_{\rm{H}_2}$}\xspace}

\def\rl10{\mbox{$R_{\rm{L10}}$}\xspace}
\def\rp06{\mbox{$R_{\rm{P06}}$}\xspace}

\def\kms{\mbox{km~s$^{-1}$}\xspace}

\def\h2{\mbox{{\sc H}$_2$}\xspace}


%% file: divisions/thesisfront.tex
\pagestyle{empty}

\ThisCenterWallPaper{1.04}{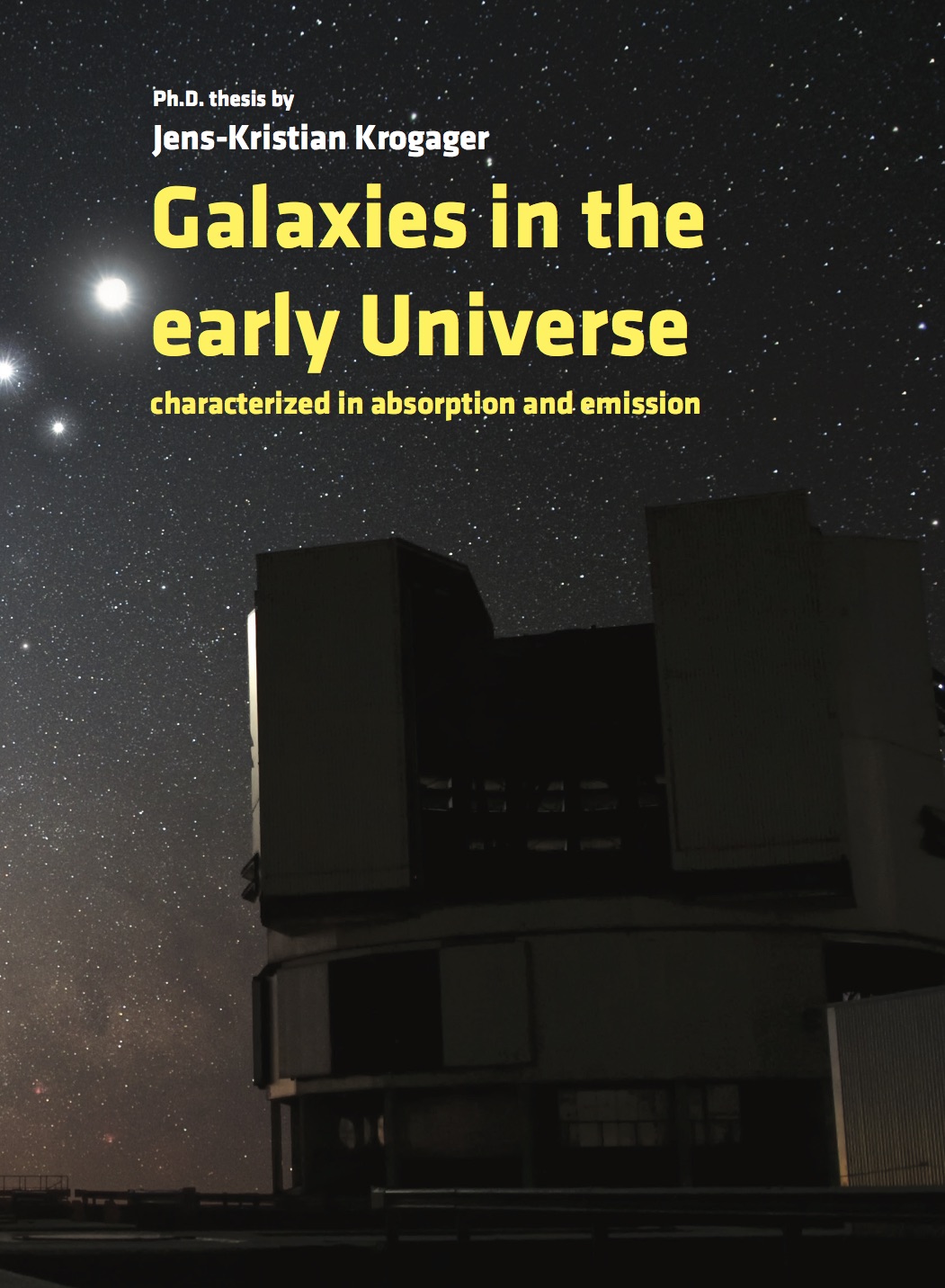}
{.}

\newpage
\clearpage

\pagestyle{empty}

\definecolor{KUgray}{rgb}{0.4, 0.4, 0.4}

\vspace*{-1.0cm}
\begin{table}
\color{KUgray}\sc
    \begin{tabular}{lr}
        {\Large university of copenhagen} \hspace{30mm} & {\Large vniversitatis hafniensis} \\
        {\large faculty of science}        &  {\large facvltatis natvralis}  \\        
    \end{tabular}
\end{table}
\vspace*{2cm}

\vspace*{1cm}
\begin{center}
\colorlet{myblue}{Black}
{\color{myblue}\Huge \scshape Galaxies in the Early Universe}\\[2mm]
{{\color{myblue} \LARGE  					
characterized in absorption and emission
}}\\[3mm]

\vspace{2cm}

\color{myblue}{\large Dissertation submitted for the degree of\\[3mm]
{\bf\Huge\scshape Philosophi\ae{} Doctor}}\\[3mm]
\color{myblue}{\large to the PhD School of The Faculty of Science, University of Copenhagen}\\[5mm]
\color{myblue}{on August 28 2015, by}\\[2mm]
\color{myblue}{\Large{\bf Jens-Kristian Krogager}}\\[1cm]
\color{myblue}{\large Supervisors:} {\large{\it Johan P. U. Fynbo} and {\it C\'edric Ledoux}} \\[10mm]

\color{myblue}{ Opponents:}\\[2mm]

{\it {Prof. Steve Warren},\\
\small Imperial College London, United Kingdom}\\[2mm]

{\it {Valentina D'Odorico},\\
\small National Institute of Astrophysics, Italy}

\end{center}

\leavevmode
\vspace{10mm}

{\small

\noindent {\it Cover design by:}\\
\indent            Snorre Rennesund\\

\noindent {\it Cover art:}\\
{\fontsize{10}{10}\selectfont
\indent            Conjunction of the Moon, Venus and Jupiter\\
\indent            over ESO’s Very Large Telescope\\
\indent            at Paranal observatory in Chile.}\\
\indent            ESO/Y. Beletsky\\
}

%% file: divisions/abstract.tex
\chapter*{Abstract}

Understanding how galaxies evolved from the early Universe through cosmic time is a fundamental part of modern astrophysics. In order to study this evolution it is important to sample the galaxies at various times in a consistent way through time. In regular luminosity selected samples, our analyses are biased towards the brightest galaxies at all times (as these are easier to observe and identify). A complementary method relies on the absorption imprint from neutral gas in galaxies, the so-called damped Ly$\alpha$ absorbers (DLAs) seen towards distant bright objects.
This thesis seeks to understand how the absorption selected galaxies relate to the emission selected galaxies by identifying the faint glow from the absorbing galaxies at redshift $z\approx2$.

In Chapters 2 and 3, the emission properties of DLAs are studied in detail using state-of-the-art instrumentation. The specific DLA studied in Chapter 3 is found to be a young, star-forming galaxy with evidence for strong outflows of gas. This suggests that the more evolved and metal-rich DLAs overlap with the faint end of the luminosity selected galaxies in terms of mass, metallicity, star formation rate, and age.
DLAs are generally observed to have low dust content, however, indications of significant reddening caused by foreground absorbers have been observed. Since most quasar samples, from which the samples of DLAs are composed, are selected through optical criteria in large all-sky surveys, e.g., Sloan Digital Sky Survey (SDSS), there might exist a bias against dusty foreground absorbers due to the reddening causing the background quasars to appear star-like in their optical colours. In Chapters 4 and 5, these hypothesized dusty absorbers are sought for through a combination of optical and near-infrared colour criteria. While a large number of previously unknown quasars are identified, only a handful of absorbers are identified in the two surveys (a total of 217 targets were observed, 137 are previously unknown).
One of these targets, quasar J2225+0527, is followed up in detail with spectroscopy from the X-shooter intrument at the Very Large Telescope. The analysis of J2225+0527 is presented in Chapter 6. The dust reddening along the line of sight is found to be dominated by dust in the metal-rich foreground DLA. Moreover, the absorbing gas has a high content of dense, cold and molecular gas with a projected area smaller than the background emitting region of the broad emission lines.

In the last Chapter, a study of the more evolved, massive galaxies is presented. These galaxies are observed to be a factor of $2-6$ times smaller than local galaxies of similar masses. A new spectroscopically selected sample is presented and the increased precision of the redshifts allows a more detailed measurement of the scatter in the mass--size relation. The size evolution of massive, quiescent galaxies is modelled by a ``dilution'' scenario, in which progressively larger galaxies at later times are added to the population of denser galaxies, causing an increase of the mean size of the population. This model describes the evolution of both sizes and number densities very well, however, the scatter in the model increases with time, contrary to the data. It is thus concluded that a combination of ``dilution'' and individual growth, e.g., through mergers, is needed.

%% file: divisions/preface.tex
\chapter*{Preface}

\vspace{-10mm}

In the beginning of the 20th century, it was realized that the Universe was not confined to our own galaxy -- the Milky Way -- but is instead made up of billions of individual galaxies like our own, the so-called `island universes'. This shift in our cosmological understanding started a wave of exploratory endeavours looking beyond the edges of the Milky Way. As we looked farther away, we made ever more puzzling discoveries: violent galactic collisions, powerful explosions of stars, and super massive black holes feeding on surrounding gas. We are now starting to see the first galaxies as they looked shortly after the Big Bang; Irregular, clumpy, and full of young stars, gas and dust, these young galaxies are far from the breathtaking spiral- and immense elliptical galaxies we know from the local Universe. An immediate question springs to mind: How did these first clumps of stars turn into the myriad of beautifully arranged galaxies that surround us? A question so complex, that an answer is not to be found in a single lifetime, let alone in a mere Ph.D. thesis. Instead of embarking on a quest so seemingly impossible, as if to play a full orchestra singlehandedly, this thesis seeks to shed light on some of the building blocks that make up part of the answer.

This thesis consists of a collection of my first-author articles written during the last five years.
The first chapters (\ref{K12} through \ref{K15}) are devoted to the study of gas in young galaxies, a matter best analysed by the use of absorption studies -- a technique through which material in front of bright objects can be studied in great detail.
In the first chapter, we will investigate how to see the unseen, that is, locating the emission from the material causing absorption towards distant, bright sources -- in this case quasars. This will in turn lead us to a search for missing quasars obscured so heavily by the foreground material that these quasars escape detection through the common classification methods. In the last chapter of the first part, we will characterize one such example of a strongly obscuring galaxy. 
 
The last chapter of the thesis (Chapter~\ref{K14}) focuses on the older and more evolved galaxies in the early Universe. These are observed to be more compact than their local analogs, and we will investigate how this evolution in size might have occurred. Rather than assuming an actual growth of each individual galaxy, we studied how the formation of larger galaxies at later times may dilute the compact population and mimic a size evolution.\\

With this short teaser in mind, we can start digging into the details of galaxy evolution...


%% file: divisions/ack.tex
\chapter*{Acknowledgements}

First of all I want to thank my PhD supervisor, Johan, for the great supervision through the last five years and for granting me the chance to pursue my childhood dreams. I am deeply grateful for the freedom that I have had during my time as a PhD student thanks to Johan. Especially my two year studentship at ESO, Chile has been an unforgettable experience. I thank all of the staff and students at ESO for all the amazing times we have shared. The list of names is too long to mention, but I owe a great debt of gratitude to Joey and Gerrit for accepting me in their home when I arrived to Chile, to Cédric for his dedication and supervision during my studentship, to Claudio (Melo) for gently pushing me to stay in science, to Claudio (Saavedra) and Pablo for moral support and for introducing me to Chilean culture, to Dany, Charles, Simona, Karla, and Sebastian, las amo a mis huevonas enfermas, and last but definitely not least to Liz and Mirjam for always making me smile no matter what! You have all been my second family through my two years in Chile.\\

I also want to thank all of my exuberant colleagues at DARK. I cannot imagine a better place to be doing research. Thank you Michelle, Julie, Corinne, Brian and Damon for your amazing support both work related and personal. Thank you Sune and Andrew for accepting me in your group; it was a pleasure working with you. Also, I thank Martin (Sparre) for assisting me through the PhD and for sharing your knowledge and great spirit, I am sure we will have many more beers in the future!\\

And of course I thank my family without whom I would never have made it through. To my parents, Pia and Ernst, and all my siblings, Ida, Johanne, Svenning, Anna, and Kathrine, thanks for accepting all my crazy work these past years. And I am deeply grateful for the support of my best friend, Pernille. I thank all my friends and my amazing choir for all the great times that helped me get my mind off work.\\

--- I owe it all to you, thank you! Gracias! Tak!

%% file: divisions/intro_DLA.tex
\chapter{Introduction}
\label{intro}

The first part of the thesis is devoted to the study of neutral hydrogen absorption systems observed towards background quasars. These come in various types depending on the strength of the Lyman $\alpha$ (\lya) absorption line. In the first five chapters (Chapters~\ref{K12} through~\ref{K15}), the focus will be on the strongest type of absorbers called damped Ly$\alpha$ absorbers. Below follows a general introduction to damped Ly$\alpha$ absorbers.
The last chapter (Chapter~\ref{K14}) of this thesis is devoted to a study of the size evolution of massive, quiescent galaxies. A short summary of the background and the definition of \textit{quiescence} will be presented later in Section~\ref{intro:evolved_gal}. Throughout the thesis, I will assume a flat $\Lambda$CDM cosmology
with $H_0=67.9\, \mathrm{km s}^{-1}\mathrm{Mpc}^{-1}$, $\Omega_{\Lambda}=0.69$ and
$\Omega_{\mathrm{M}} = 0.31$ \citep{Planck2014}.

\section{What are Quasars?}
\begin{figure}
    \centering
    \includegraphics[width=0.7\textwidth]{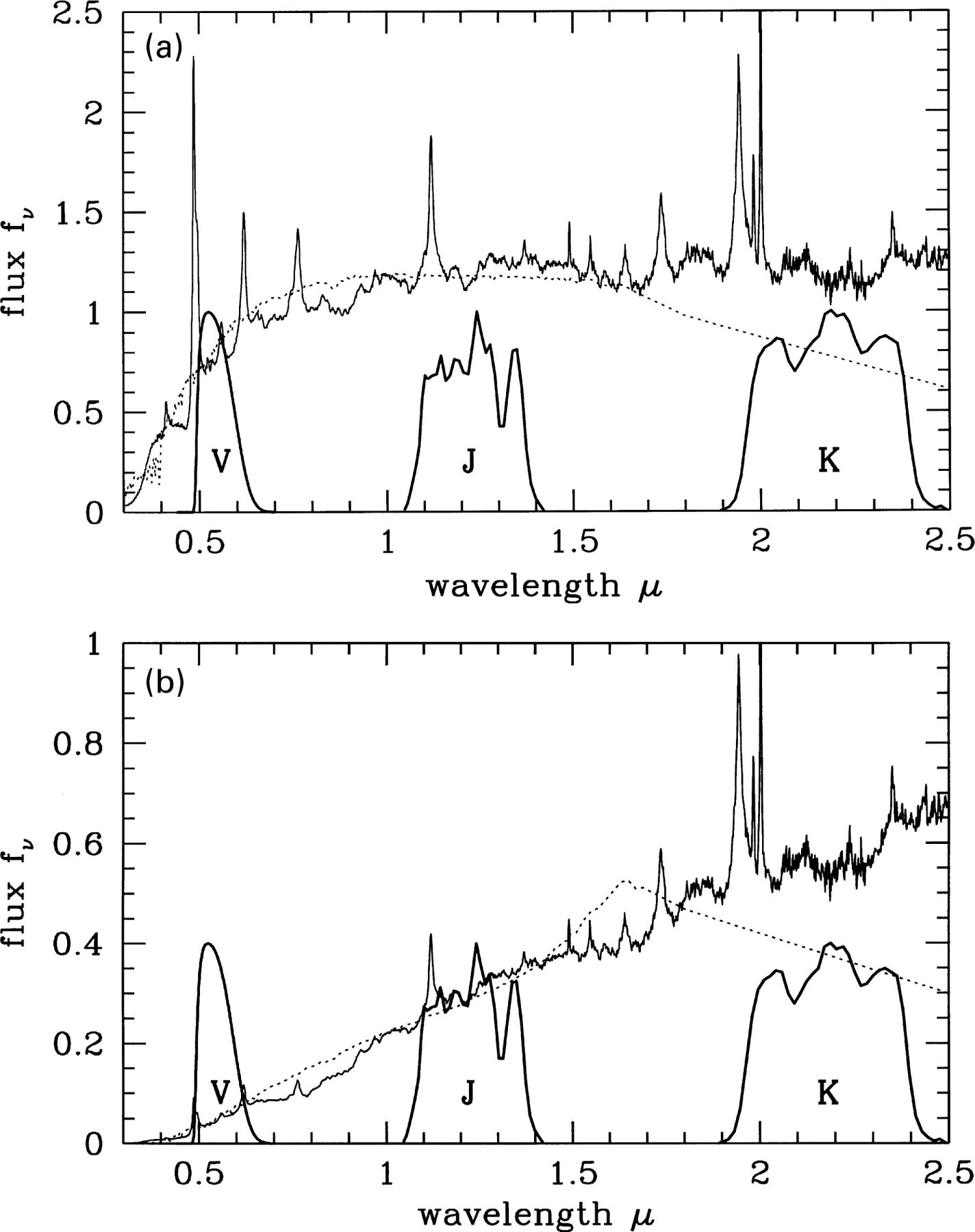}
    \caption{Comparison of the spectral energy distribution of quasars and stars, illustrating the principle of the KX method. In each panel, the quasar template is shown by the solid line, and a stellar template of similar $V-J$ colour is shown by the dotted line. The $V$, $J$ and $K$ transmission curves are over-plotted.
    The upper plot shows the spectra of an unreddened $z=3$ quasar and an early K star. The lower plot shows the same quasar template reddened by $E(B–V)=0.3$ in an intervening system at $z=2.5$. The spectrum of an early M star is shown for comparison. The $K$-band excess is clearly visible in both panels, demonstrating the power of the KX method in identifying both reddened and unreddened quasars. This figure is adapted from \citet{Warren2000}.}
    \label{fig:KX}
\end{figure}

Quasi-stellar objects (QSOs) -- or quasars -- were first detected as radio sources appearing star-like in their optical counterparts \citep{Schmidt1963, Matthews1963, Hazard1964}. These intrinsically very luminous objects are now linked to the accretion process of gas onto a massive black hole in the centres of galaxies. Quasars are characterised by a very strong energy output at almost every wavelength from X-rays to radio, and at optical wavelengths the quasars show very characteristic, broad emission lines (Type I). Owing to their high luminosity over a large range of wavelengths, quasars can be selected in many ways, typically through radio or X-ray surveys, or in large optical all-sky surveys such as the Sloan Digital Sky Survey \citep{York2000} or the 2dF QSO redshift survey \citep{Croom2004}. The quasar selection in optical surveys is primarily based on the fact that quasars appear more blue than stars \citep[the so-called ultraviolet excess, UVX; e.g.,][]{Sandage1965, Schmidt1983, Marshall1984}. However, such optical criteria are only efficient out to redshift $z\approx2.2$ where the \lya\ forest starts to enter the optical blue filters. Moreover, the optical criteria are more susceptible to the effects of dust as they probe the rest-frame UV of the quasar where the reddening from dust is strongest. For these reasons, a selection method at larger wavelengths was introduced. The technical progress in near-infrared instrumentation lead to the development of the KX method, which utilizes the excess of quasars relative to stars in the $K$-band \citep{Warren2000}. The main principle of the KX method is demonstrated in Figure~\ref{fig:KX}. 

Recently, more advanced selection techniques are being used to select large samples of quasars based on optical photometry, e.g., neural networks and the so-called extreme deconvolution \citep[see][and references therein]{Ross2012}. Many of these methods obtain not only the classification, but also an estimate of the photometric redshift, $z_{\rm phot}$.
With the onset of large, high-cadence sky surveys, one can also use the fact that quasars exhibit temporal variations in brightness to select candidates based on their light curves \citep{Schmidt2010, Graham2014}. Selecting quasars on the basis of their variability will allow for an efficient selection, which is not biased by dust in the quasar nor in intervening systems.
Furthermore, the extreme astrometric precision ($\sim300~\mu$as~yr$^{-1}$ in proper motion) of the {\it Gaia} mission \citep{deBruijne2012} allows an unbiased selection of quasars due to the fact that quasars have zero proper motion. For more details about this selection method, see \citet{Heintz2015}. 

The quasars used throughout the thesis are optically selected quasars from the Sloan Digital Sky Survey (SDSS). In Chapters~\ref{redQSOs} and \ref{HAQ}, we investigate a possible bias in the optically selected quasar samples by employing colour criteria in the optical and near-infrared -- these criteria are similar to the KX method described above.

\section{What are damped Ly$\alpha$ absorbers?}
\label{intro:DLA_general}
\begin{figure}
    \centering
	\includegraphics[width=0.95\textwidth]{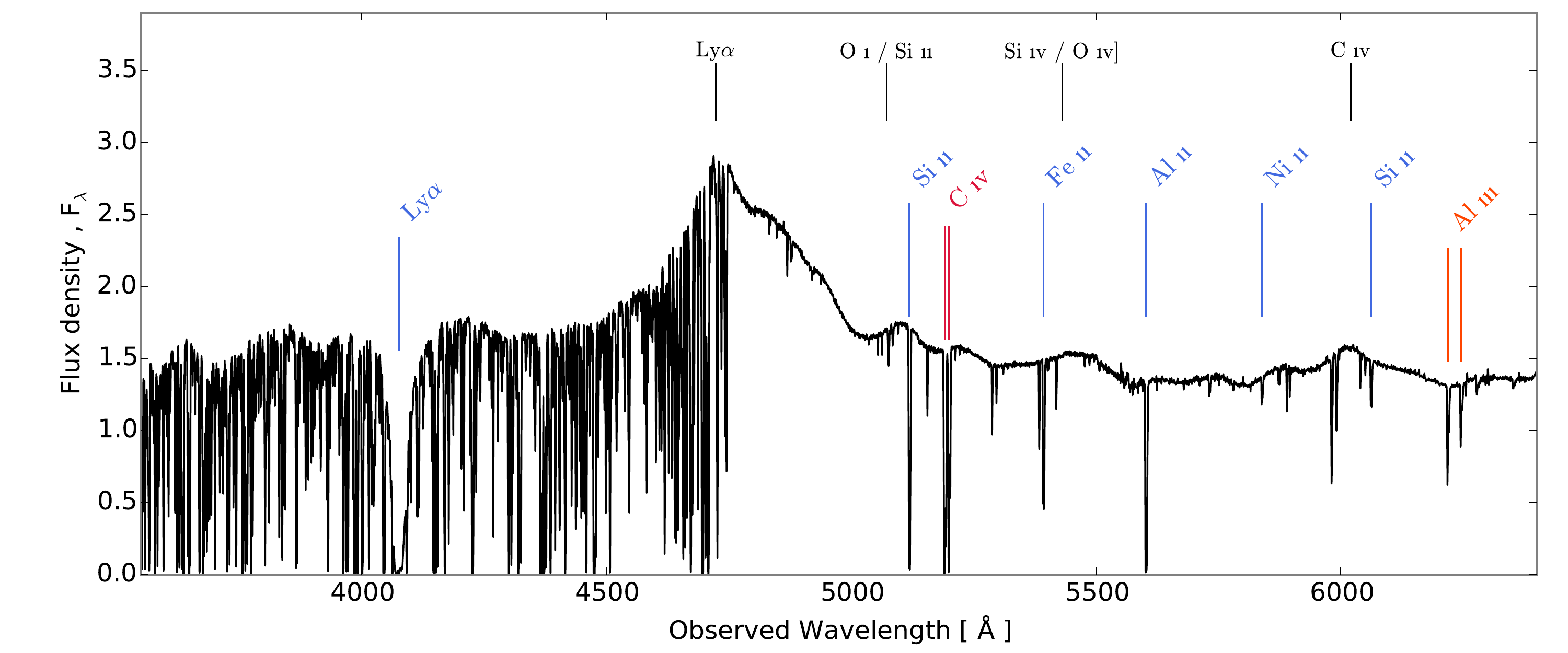}
	\caption{Spectrum of a quasar at $z=2.89$ with an intervening DLA at
	$z=2.35$. The strong, damped absorption line from \lya\ is clearly
	seen in the Ly$\alpha$ forest. The most prominent broad emission lines
	from the quasar are marked in black, and the strongest low-ionization
	metal lines from the DLA are marked in blue. The higher ionization
	lines from the DLA are marked in orange and red.}
	\label{fig:QSO-DLA}
\end{figure}

For decades, damped Lyman-$\alpha$ absorbers (DLAs) have been used to study the cosmic reservoir of neutral gas from redshift $z\approx 1.6$ out to a redshift of $z\approx5$ \citep{Noterdaeme2012b, Rafelski2014}. Rather than a physical cut, the lower limit on redshift is defined for ground-based surveys that are restricted by the atmospheric cut-off at $3200$~{\AA}. DLAs are observable at lower redshifts only from space. 
As mentioned above, damped Lyman-$\alpha$ absorbers are the strongest absorption features in the family of neutral hydrogen absorbers.
In Figure~\ref{fig:QSO-DLA}, a typical quasar spectrum with an intervening DLA is shown.
The strong absorption is clearly visible in the dense collection of weaker Ly$\alpha$ absorption lines, the so-called Ly$\alpha$ forest.

DLAs are defined as having column densities in excess of $\NHI>2\times10^{20}~\rm{cm}^{-2}$.
Although appearing rather arbitrary, this definition ensures a high neutrality of the gas in DLAs, i.e., the gas is almost entirely self-shielding contrary to other Ly$\alpha$ absorbers (Ly$\alpha$ forest and Lyman-limit systems).
The high neutral gas fraction in DLAs means that ionization corrections are negligible when determining metal line abundances. For this reason it is valid to assume that the singly ionized metal lines\footnote{Throughout the thesis, the ionization state of a given element is marked by small roman numerals, e.g., the neutral state is referred to by {\sc i}, the first ionized state is marked by {\sc ii}, and so forth.} trace the total metal abundance, that is: $N({\rm Fe}) = N\left({\rm Fe\,\textsc{ii}} \right)$ and so on.
Therefore, DLAs serve as some of the most precise probes of chemical enrichment at high redshift.

The first large sample of DLAs was published by \citet{Wolfe1986}, who were looking for high-redshift gas-rich disc galaxies by searching for their Ly$\alpha$ absorption in quasar spectra. Since then, many surveys have been carried out \citep{Lanzetta1991, Lanzetta1995a, Wolfe1995, Storrie-Lombardi2000, Ellison2001a, Peroux2003b}, and especially the spectroscopic database of the Sloan Digital Sky Survey (SDSS) has increased the sample sizes significantly \citep{Prochaska2003a, Noterdaeme2009b, Noterdaeme2012b}.
One of the main advantages of the absorption selection technique is that the galaxies are not selected based on their luminosity, but rather on their absorption cross-section. That way a more representative sample of the luminosity function is achieved \citep{Wolfe1986}.

\begin{figure}
    \centering
	\includegraphics[width=0.6\textwidth]{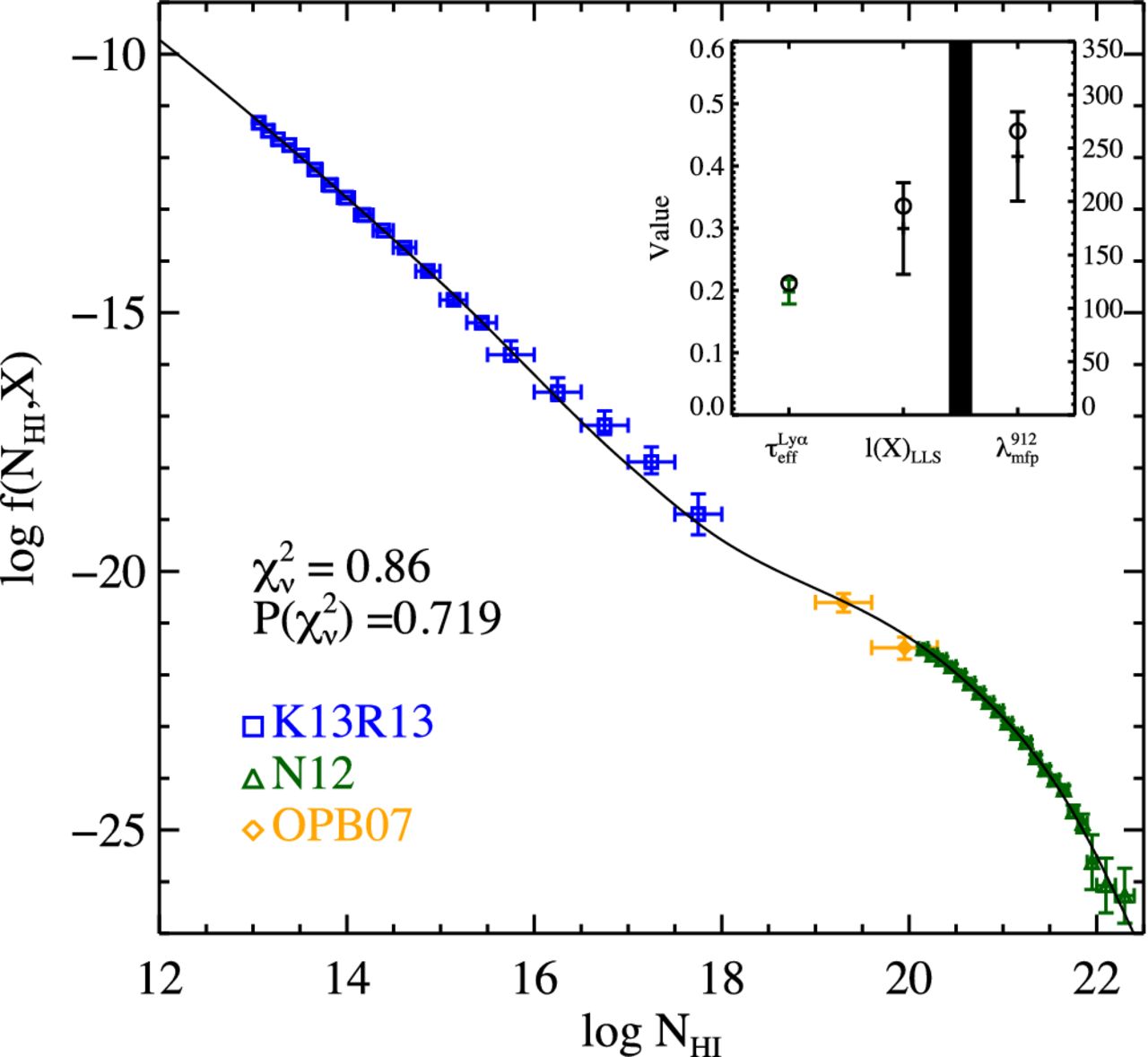}
	\caption{Column density distribution function of quasar absorbers normalized by the absorption redshift path. The data shown are a collection from \citet[][K13]{Kim2013}, \citet[][R13]{Rudie2013}, \citet[][N12]{Noterdaeme2012b}, and \citet[][OBP07]{oMeara2007}. The figure is taken from \citet[][their fig.~7]{Prochaska2014}.}
	\label{fig:fNHI}
\end{figure}

Figure~\ref{fig:fNHI} shows the distribution function of neutral hydrogen absorbers as function of the column density of \ion{H}{i}, $f(\NHI, X)$, from \citet{Prochaska2014}. The blue points mark \lya\ forest absorbers, the yellow points show the so-called super Lyman-limit systems (also sometimes referred to as sub-DLAs), and the green points mark the DLAs. The normalizing quantity, $X$, represents the absorption path along a given line-of-sight out to redshift, $z$, and is defined as:
$$ X(z) = \int_0^z \frac{H_0}{H(z')} (1+z')^2\ dz'~,$$
\noindent
where $H_0$ is the Hubble constant, and $H(z)$ denotes the Hubble parameter which for a flat $\Lambda$CDM universe is $H(z) = H_0\ \sqrt{(1+z)^3 \Omega_{\mathrm{M}} + \Omega_{\Lambda}}$.
By integrating over $f(\NHI, X)$, one can obtain the mass density of neutral gas:

$$ \Omega_g(X)dX = \frac{H_0\ \mu\ m_H}{c\ \rho_c} \int_{N_{\rm min}}^{N_{\rm max}} \NHI f(\NHI, X)\ d\NHI\ dX~,$$
\noindent
where $\mu$ is the mean molecular weight ($\mu=1.3$); $m_H$ is the mass of a hydrogen atom, and $\rho_c$ is the current critical density. If one sets the lower integration boundary to the limiting column density for DLAs ($N_{\rm min}=2\times 10^{20}~cm^{-2}$) and integrates over all larger $\NHI$ ($N_{\rm max}=\infty$), one obtains a measure of the contribution from DLAs to the total gas mass density.
Due to the shape of the distribution function, damped \lya\ absorbers dominate the neutral gas content in the Universe for all redshifts lower than $z\approx5$ \citep*{Noterdaeme2009b}. And as such, DLAs are important tools in our understanding of galaxy formation and evolution since they probe the neutral gas that eventually may condense and form stars. 

\newpage

\subsection{Metal Absorption in DLAs}
\label{intro:metals}
Along with the strong \lya\ absorption line seen in the quasar spectrum we always observe corresponding absorption lines from metal transitions, the dominant lines being from low-ionization lines. Higher ionization states are also typically observed, e.g., \ion{C}{iv}, \ion{Al}{iii}, and \ion{O}{vi}, indicating that a DLA sightline traces multiple interfaces of gas clouds. The low-ionization metal lines typically show very complex line structures with many components, especially at high metallicities \citep{Prochaska1997}. There are several ways to recover the column densities of the metal absorption lines \citep[e.g.,][and references therein]{Savage1991}; However, in this thesis I have chosen to fit the line profiles as this allows a better handling of blended and contaminated lines. For this purpose, the most appropriate method is to fit strong, though not saturated, absorption lines with a Voigt profile (saturated lines are defined later in Sect.~\ref{intro:data}). The Voigt profile is obtained by convolution of the Gaussian and the Lorentzian distribution functions; However, for practical purposes, an analytical approximation is used. An absorption line is defined by the column density of the given species, $N$, the broadening parameter, $b$, and a set of atomic parameters for the specific transition (here marked by subscript $i$): the resonant wavelength, $\lambda_i$; the oscillator strength, $f_i$; and the damping constant, $\Gamma_i$. When fitting the absorption lines, one seeks to recover $N$ and $b$, assuming that the atomic parameters are well-determined. The details of the fitting method are presented in Appendix~\ref{appendix:Voigt}.

\subsubsection{Probing the nature of DLAs from metal absorption lines}
\begin{figure}
    \centering
    \includegraphics[width=0.7\textwidth]{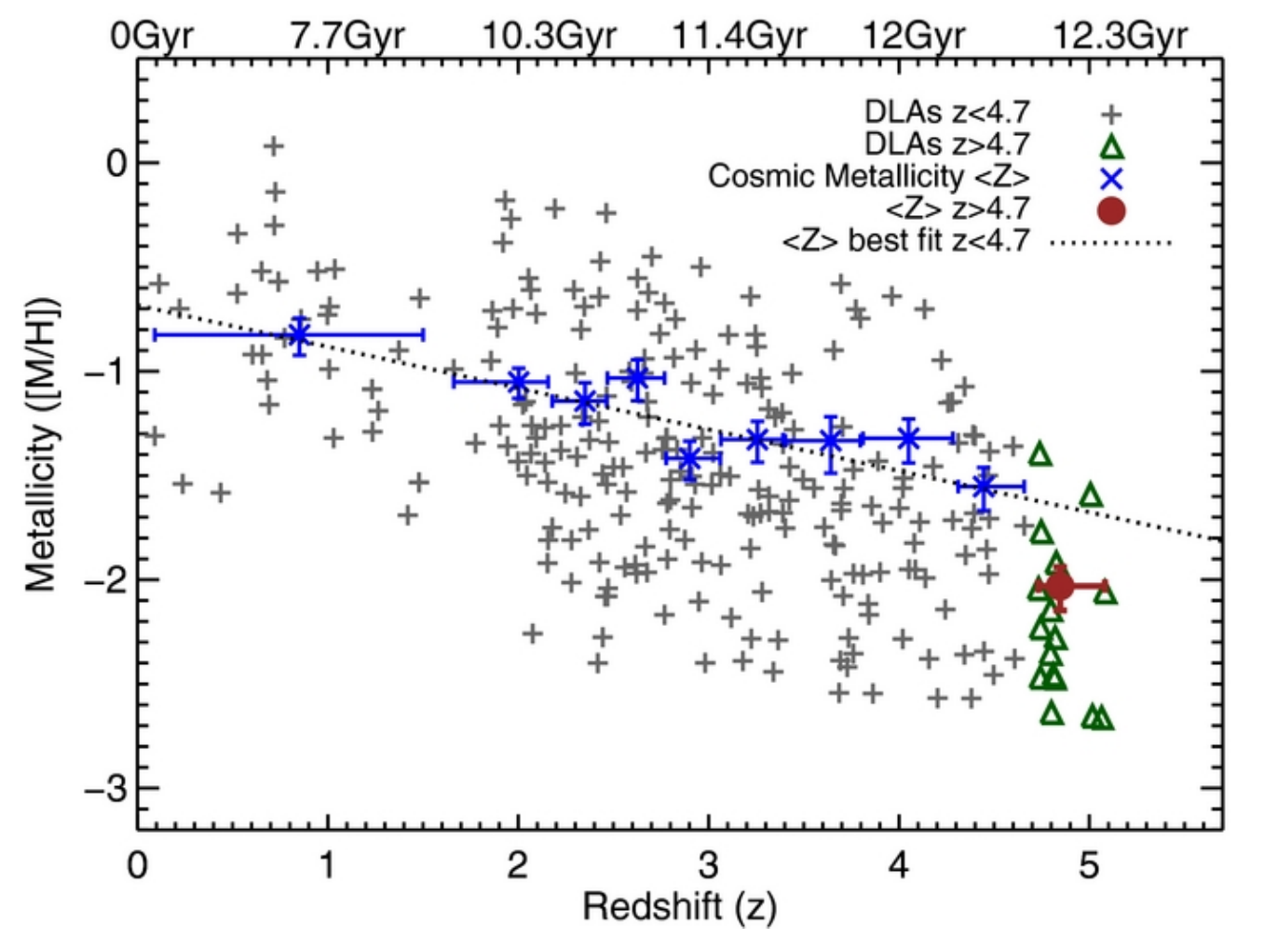}
    \caption{Cosmic metallicity evolution of DLAs as function of redshift.
    From \citet[][their fig.~3]{Rafelski2014}.}
    \label{fig:metal_evolution}
\end{figure}

By measuring the abundances of metals present in DLAs, e.g., S, Si, Fe, Zn, it is possible to map out the chemical enrichment in the densest regions throughout most of cosmic time\footnote{In the assumed cosmology, a redshift of $z=5$ corresponds to a look-back time of 12.62~Gyr, i.e., around 90 per cent of the age of the Universe.} \citep[][]{Kulkarni2002, Prochaska2003b, Rao2005, Rafelski2014}. Figure~\ref{fig:metal_evolution} shows the latest compilation of metallicities, $[{\rm M/H}]$\footnote{Throughout the thesis I will use the notation $[{\rm M/H}]$ to denote the abundance ratio of an arbitrary indicator, usually a volatile element such as Zn or S, relative to the Solar abundance ratio.}, as a function of redshift from \citet{Rafelski2014}.
The study presented by \citeauthor{Rafelski2014} shows that DLAs on average are metal poor at all times relative to Solar and that the average metallicity (blue points) shows a gradual increase. If the relation is extrapolated to $z=0$, a local value of $Z_0/Z_{\odot} \approx 0.2$ is found. However, the overall population of disc galaxies locally have metallicities around $Z/Z_{\odot}\sim1$. This apparent metal deficiency led to the conclusion that DLAs could not be the progenitors of today's star-forming disc galaxies \citep[e.g.,][]{Pettini1999}.

One of the hypotheses back when the first DLA samples were being assembled was that galaxies formed from a single collapse of a massive gas cloud, which would lead to the formation of a gaseous, thin disc \citep[e.g.,][]{Eggen1962}. The \lya\ absorption seen towards quasars was thought to trace these gas discs of galaxies in the making. This was further studied through the kinematical information recovered from the velocity structure of the metal lines.
\citet{Prochaska1997} (and also \citealt{Ledoux1998}) use the velocity width, \dv, of the low-ionization lines to investigate the underlying population of galaxies responsible for the absorption. These authors define \dv as the velocity encompassing 90 percent of the apparent optical depth of the line, $\uptau' = -\ln(I_{\rm obs})$.

From the distribution of $\uptau'$ for a given line, one can calculate the line centroid, $\lambda_0$, along with the 5$^{\rm th}$ and 95$^{\rm th}$ percentiles, $\lambda_5$ and $\lambda_{95}$, respectively. The velocity width is then obtained as follows:
$$\Delta V_{90} = c\, \frac{\lambda_{95} - \lambda_{5}} {\lambda_0}~.$$
\noindent
By modelling absorption lines through various geometries and quantifying these modelled spectra in terms of \dv and the component structure, Prochaska et al. conclude that the observed absorption line profiles are consistent with those expected from large rotating discs of gas \citep*{Prochaska1997, Prochaska1998}.
However, within the $\Lambda$CDM paradigm of cosmology, galaxies are predicted to form via hierarchical merging of substructures; hence high-redshift DLAs should originate in these galactic clumps rather than being signatures of large disc galaxies \citep[e.g.,][]{Tyson1988, Gardner1997, Haehnelt1998}.

How do we connect our knowledge about the gas phase absorption to the galaxies we observe in direct emission? Chapters~\ref{K12} and \ref{K13} are devoted to the study of this puzzling link between absorption and emission properties of the galaxies causing DLAs.

\subsection{Dust and Molecules in DLAs}
\label{intro:dust}
Dust is a ubiquitous substance throughout the Universe and its presence is linked to many phases of the evolution of stars, galaxies and quasars. Nevertheless, most astronomers have a peculiar love-hate relationship with cosmic dust.
On the one hand, dust properties are important and indeed very interesting to study as the dust grains serve as primary catalysts for the formation of molecular hydrogen in the interstellar medium \citep[e.g.,][]{Hollenbach1971}. Moreover, the dust grains in the early stellar phases clump together to produce planetesimals and subsequently facilitate the formation of life \citep[e.g.,][]{Lissauer1993}.
On the other hand, dust scatters the light from distant objects hampering their detection and analysis.

\subsubsection{The observable effect of dust}
The main observable effect of dust grains is the scattering of incident light rays. This results in two effects: it makes the source appear fainter and redder, due to the wavelength dependent scattering. This is most commonly quantified in terms of a `reddening curve' (or extinction law/curve), which ascribes a functional form to the wavelength dependent attenuation $A(\lambda)$ relative to the attenuation in the optical $V$-band, $A(V)$.
This is also sometimes given in terms of the `colour excess', $E(B-V) = A(B)-A(V)$, i.e., the difference between the observed colour $(B-V)$ and the intrinsic colour $(B-V)_0$. The colour excess and attenuation are linked through the parameter $R_V \equiv A(V)/E(B-V)$, which simply put is a measure of the slope of the reddening curve. Various functional forms exists; however, the main distinction can be made between Milky Way-type and Small Magellanic Cloud (SMC) type depending on the presence or absence, respectively, of the so-called $2175$~\AA\ bump. The exact origin of this feature is not known, but it is usually ascribed to carbonaceous dust grains \citep[e.g.,][and recently Misma \& Li, 2015]{Draine2003}\nocite{Misma_arxiv}. A general functional form used to describe reddening curves is the \citet{FM1990} formulation\footnote{Here I give the expanded formalism including the $c_5$ parameter introduced by \citet{FM2007}}, which provides an analytical form with 7 parameters:

\begin{equation}
    k(\lambda-V) \equiv  \frac{E(\lambda-V)}{E(B-V)} =
    \begin{cases}
        c_1 + c_2x + c_3 D(x, x_0, \gamma), & x\le c_5 \\
        c_1 + c_2x + c_3 D(x, x_0, \gamma) + c_4 (x-c_5)^2, & x>c_5 
    \end{cases}
\end{equation}
\noindent
where $x=\lambda^{-1}$ is the reciprocal wavelength and $D(x,x_0,\gamma)$ is the broad Drude component to represent the $2175$~\AA\ bump:

\begin{equation}
    D(x, x_0, \gamma) = \frac{x^2} {(x^2-x_0^2)^2 + x^2\gamma^2}~,
\end{equation}
\noindent
for which the peak location and width are given by $x_0$ and $\gamma$, respectively.
In the following Chapters, I will use the extinction curve in terms of \Av rather than $E(B-V)$:

\begin{equation}
    \xi(\lambda) = \frac{A(\lambda)}{A(V)} = \frac{1}{R_V}k(\lambda-V) + 1~.
\end{equation}

\noindent
In Chapter~\ref{redQSOs}, we used the extinction curves parametrized by \citet{Pei1992}; whereas in Chapters~\ref{HAQ} and \ref{K15}, we used the best-fitting parameters from \citet{Gordon2003}, who obtain parameters for Small Magellanic Cloud (SMC), Large Magellanic Cloud (LMC), and Milky Way (MW) types.

\subsubsection{Dust in DLAs}
If DLAs are, as hypothesized, associated to the gas in star-forming galaxies then one would expect the DLAs to harbour dust grains, as dust is expected both in the formation process of stars and during the final stages of stellar evolution. Therefore, several studies have searched for dust in DLAs \citep[e.g.,][]{Fall1989, Pei1991}. The way to look for dust in a DLA is usually to look at the reddening effect of the dust on the background quasar. This can either be done in a statistical approach, where one compares samples of quasars with and without DLAs, or in an individual object.

\newpage

The statistical approaches can be divided into three groups:

\begin{enumerate}
    \item {The first approach is based on spectroscopic data where the reddening may be recovered by fitting the slope of the quasar spectrum in the rest-frame UV \citep{Pei1991, Murphy2004, Kaplan2010}.}

    \item {The second approach is purely photometric in nature and relies on the measurement of colours of the quasar \citep[e.g.,][]{Maiolino2001b, Vladilo2008, Maddox2008, Khare2012}.}
\end{enumerate}

\noindent
In both the first and second approaches, the reddening is estimated by comparing the metric (either UV slope or colour) for the sample of quasars with and without DLAs. This way the reddening can be estimated with no assumption regarding the extinction curve. 

\begin{enumerate}
    \setcounter{enumi}{2}
    \item {The third approach uses stacking to infer an average spectrum of quasars with and without DLAs. The ratio of these two stacked spectra will reveal the average wavelength dependent attenuation \citep{Frank2010, Khare2012}. This approach is therefore sensitive to the type of extinction curve.}
\end{enumerate}

While these approaches work well on large statistical samples, they fall short when trying to estimate the amount of dust in any single DLA. Instead, the presence of dust in a DLA can be inferred purely by looking for the dust emission from the DLA galaxy and by analyzing the abundance ratio of metals \citep[e.g.,]{Khare2004, Vladilo2005, Vladilo2008}. Elements such as iron, chrome, \&c. (the so-called refractory elements) have a high affinity for dust. As the refractory elements condensate and form dust, the gas-phase abundance of these elements will decrease compared to non-refractory (volatile) elements, e.g., Zn or S. The commonly used ratio [Fe/Zn] thus indicates how much iron has been depleted from the gas-phase -- assuming an intrinsic abundance pattern of the DLA. The exact intrinsic abundances for DLAs has been studied in great detail, however, no consensus has yet been reached \citep[for a recent discussion, see][and also \citealt{Rafelski2012}]{Berg2015}.
The depletion can be translated to a measure of the attenuation through the theoretical relation between attenuation and the column density of iron in the dust phase \citep[e.g.,][]{Vladilo2006, DeCia2013}. This method as well as its caveats are described in more detail in Chapter~\ref{K15}. 

Last but not least, one can use quasar templates (quasar spectra stacked to generate a mean spectrum) to fit the observed spectrum with an assumed dust model. This is the approach which is used in Chapters~\ref{redQSOs}, \ref{HAQ}, and \ref{K15} to infer \Av towards quasar sightlines. The model is presented in detail in Appendix~\ref{appendix:dust_model}.

In general, the effect of dust in existing DLA samples is not large, and some studies completely dismiss any significant dust reddening from DLAs \citep{Murphy2004, Frank10}. Nonetheless, the DLAs with higher metallicities are found to have higher depletion ratios ([Fe/Zn]) and thus probably higher dust content \citep{Ledoux2003}. In later studies, it was shown that if the analysis is limited to the metal-rich DLAs, a significantly higher amount of dust is inferred \citep{Kaplan2010, Khare2012}. The subsequent detection of individual DLAs with high \Av \citep{Fynbo2011, Noterdaeme2012a, Wang2012} led us to the search for dust reddened quasars. The hypothesis is that metal-rich and dusty DLAs might be missed in the optically selected quasar samples due to the reddening effects of dust in the foreground absorber \citep{Fall1989, Pei1991, Pei1995, Pontzen2009}. The current samples of DLAs would therefore be biased against systems with large dust column densities, high metallicities, and typically also high $\NHI$ \citep[see also][]{Boisse1998}. Following this hypothesis, we initiated a spectroscopic survey for red quasars, based on optical and near-infrared photometric criteria. The first pilot study is presented in Chapter~\ref{redQSOs}. Building on the experience of the pilot study, the spectroscopic survey was expanded and revised. In Chapter~\ref{HAQ}, the revised criteria and results are presented.

\subsubsection{Molecular hydrogen absorption in DLAs}
Assuming that the gas in DLAs are related to {\it in-situ} star formation, the presence of H$_2$ would be expected in DLA sight lines. However, detections of H$_2$ are very scarce.
The first detection of H$_2$ was reported by \citet{Levshakov1985} and \citet{Foltz1988}, and since then very few systems with molecular hydrogen have been detected \citep[see][and references therein]{Balashev2015}.

From the few detections at hand, a low molecular fraction, $f_{\rm H_2} \lesssim 0.01$, is usually inferred \citep{Ledoux2003, Noterdaeme2008, Srianand2008a, Noterdaeme2010, Jorgenson2014, Noterdaeme2015b}. This is about a factor of 10 less than for sightlines through the interstellar medium (ISM) of the Milky Way.
The low molecular fraction in DLAs is commonly ascribed to the low metallicities in the average population.
As mentioned above, dust grains are important catalysts for the formation of H$_2$ and the amount of dust is related to the amount of metals in the gas phase.
Therefore, the low metallicities in DLAs result in low efficiencies for H$_2$ formation. Also, the Milky Way sightlines probe the interstellar material in a different way than the DLA sightlines do. A DLA sightline will typically intersect multiple phases of the ISM of a galaxy, whereas the Milky Way sightlines usually only probe a single cloud \citep{Noterdaeme2015b}. The hydrogen column density for DLAs will therefore trace the entire multiphase medium, and not just the local medium around the molecular phase.

Furthermore, as we will see in Chapters~\ref{redQSOs}, \ref{HAQ}, and \ref{K15}, the dust-rich DLAs, which are more likely to have strong H$_2$ absorption, might be underrepresented in current samples due to the dust obscuration bias against dusty sightlines mentioned above.

\section{X-Shooter Data}
\label{intro:data}
A substantial part of this thesis is based on spectra obtained with the X-shooter spectrograph mounted on the Very Large Telescope operated by ESO at Paranal, Chile.
The instrument splits the incoming light into three separate spectrographs (called arms: UVB, VIS, and NIR) thereby covering the full range of wavelengths from the atmospheric cut-off to the $K$-band simultaneously. The three arms operate at medium resolution ranging from $\mathcal{R}\sim3000-10,000$ in UVB, $\mathcal{R}\sim5000-18,000$ in VIS, and $\mathcal{R}\sim4000-11,000$ in NIR depending on the used slit-width and seeing. Further details about the data are presented in the respective chapters.

\begin{figure}
    \centering
    \includegraphics[width=1.0\textwidth]{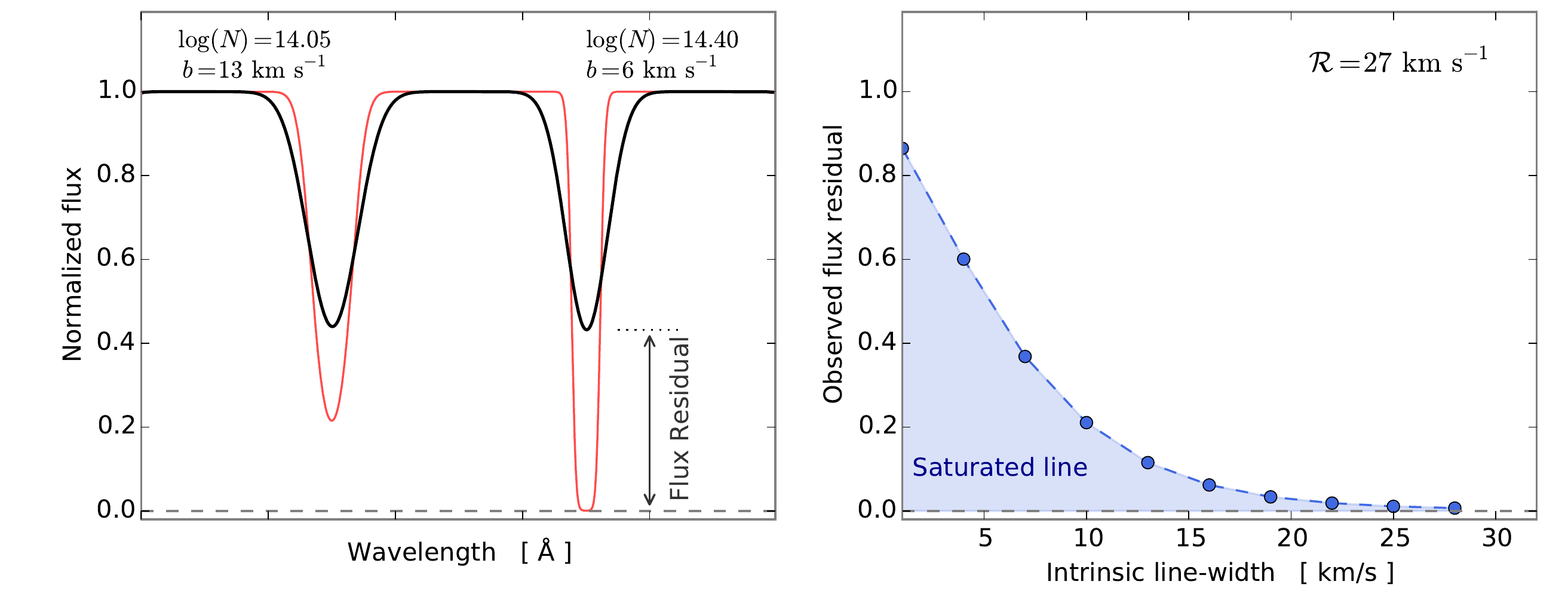}
    \caption{Simulated line profiles ({\it left}) and flux residual of
    a saturated line as function of its intrinsic line-width, $b$ ({\it right}).
    The left panel shows two absorption lines with almost identical flux residuals.
    The thin, red line shows the intrinsic line profile, and the thick,
    black line shows the same line profile convolved with the instrumental
    resolution of $\mathcal{R}=27$~\kms as it would be observed.
    For each line, the line-width, $b$, and column density, $\log(N)$,
    are given above the line. Although appearing similar in the observed
    line profile, the right absorption line is clearly saturated.
    This effect is due to the so-called hidden saturation
    caused by the instrumental line spread function, here assumed to be Gaussian.
    In the right panel, the flux residual of a saturated absorption line is shown
    as function of its intrinsic line-width. If the flux residual at peak absorption
    is below the curve, i.e., in the shaded region, the line will be saturated.
    Intrinsic line-widths for DLAs are typically larger than $b \gtrsim 4$~\kms;
    hence, a line with flux residual larger than $\sim0.6$ will be unaffected by
    hidden saturation independent of the line-width.
    Note that for very small $b$-values, this function is not well-behaved as
    numerical artefacts start to dominate the evaluation of the line-centre.}
    \label{fig:hidden_sat}
\end{figure}

For the spectral analysis in this thesis (Chapters~\ref{K13} and \ref{K15}), the spectra were obtained with a resolution of $\mathcal{R}=11,000$ in the visual arm (VIS). The spectral resolution results in a broadening of spectral features, especially for features that are intrinsically narrower than the resolution. The broadening can lead to an effect called {\it hidden saturation}, where intrinsically narrow and saturated lines appear unsaturated due to the spectral broadening of the instrument. The effect of hidden saturation is shown schematically in Figure~\ref{fig:hidden_sat}.
Hidden saturation can have severe implications for the determination of the column densities of strong metal lines, since saturated lines are not sensitive to the column density. Therefore, strong lines should be avoided when inferring column densities from spectral absorption lines.
The definition of a `strong' line depends on the instrumental resolution as well as the unknown intrinsic line width. For X-shooter working at $\mathcal{R}=11,000$ ($27$~\kms), an absorption line with flux residual at peak absorption of less than 0.6 is potentially affected by hidden saturation depending on the intrinsic line-width, $b$, see right panel of Figure~\ref{fig:hidden_sat}.

By limiting the analysis to lines that have flux residuals at peak absorption of less than 0.6, we can be fairly sure that the results will not be biased, unless the lines are extreme narrow intrinsically. Such narrow lines, however, are not observed in the low-ionization metal lines, but are more often seen in molecular H$_2$ lines.
In Chapter~\ref{K15}, we include a few transitions (\ion{Fe}{ii} and \ion{S}{ii}) which are slightly stronger than the limit derived above. However, since we also include a weaker iron transition, which is definitely not saturated, we can be sure that the fit will not be biased by any possible hidden saturation. The fit for the sulphur line might be slightly underestimated, since we cannot correct for the effect without other weaker transitions available in the spectral range.

%% file: divisions/intro_grism.tex
\section{Evolved Galaxies}
\label{intro:evolved_gal}
\begin{figure}
	\centering
	\includegraphics[width=1.0\textwidth]{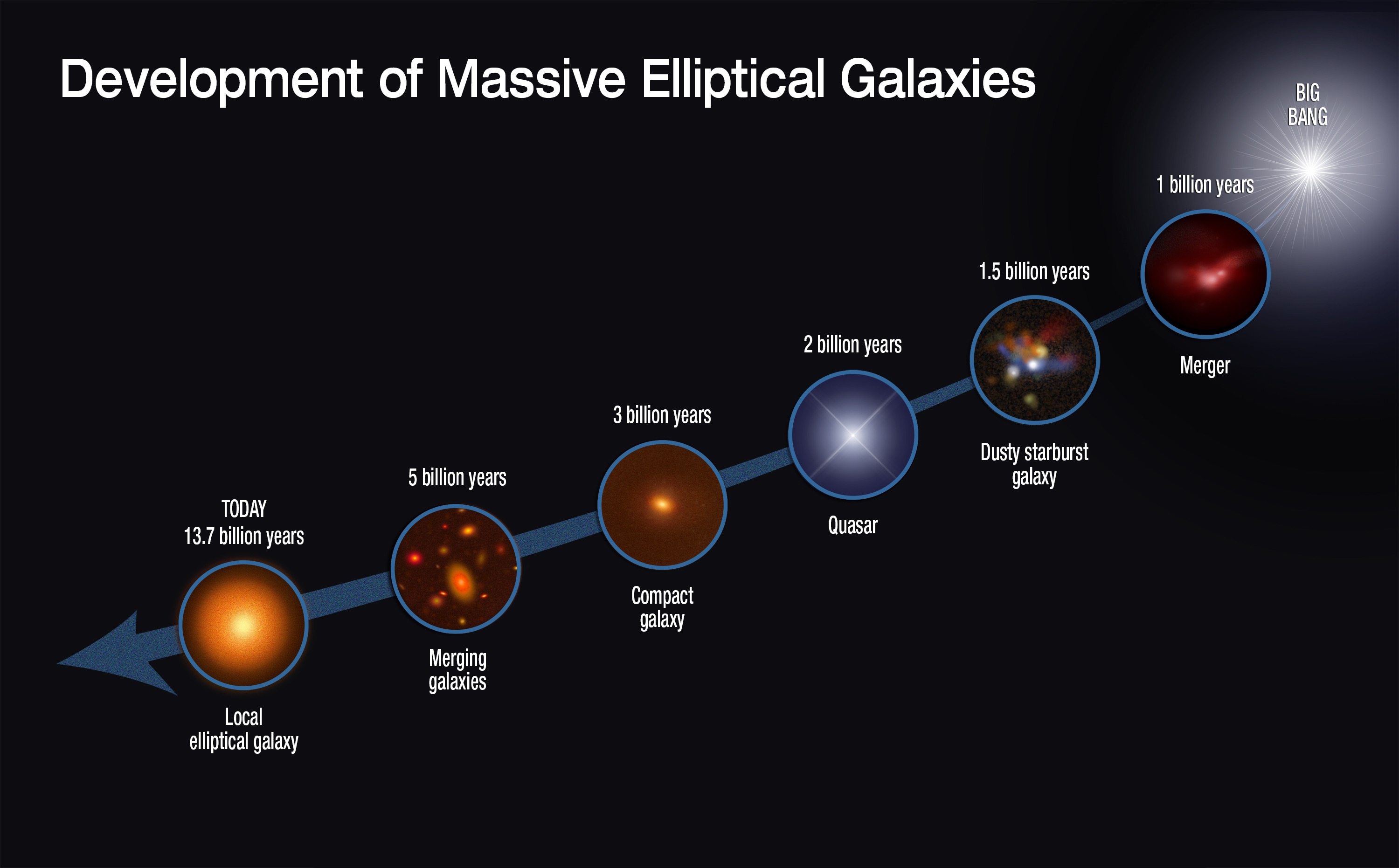}
	\caption{An illustration of the proposed evolutionary phases
    for large elliptical galaxies observed locally. The merging of gas-rich
    galaxies in the early Universe ignites a starburst, creating
    a remnant shrouded in dust. An active quasar phase quenches
    further star formation, resulting in a compact galaxy, which subsequently
    grows through merging with small, gas-poor satellite galaxies.\\
	{\small (Credit: Sune Toft and the NASA/ESA {\it Hubble Space Telescope} press office. Release number 14-028, January 2014.)} }
	\label{fig:galaxy_evolution}
\end{figure}

Thanks to the development in near-infrared (NIR) instrumentation both from ground and space we have been able to study the $z=2$ Universe in high detail. One of the most striking discoveries at high redshift is the existence of a population of old and evolved galaxies \citep{Franx2003}, the so-called distant red galaxies (DRGs).
Using high quality NIR imaging from the {\it Hubble Space Telescope (HST)}, several works have shown that these massive (${\rm M_{\star} > 10^{11}~M_{\odot}}$) galaxies are much more compact than local galaxies with similar masses.
The origin and subsequent evolution of these compact, evolved galaxies are still being investigated heavily. One of the most favoured evolutionary scenarios explains the formation of the compact galaxies through early gas-rich interactions at high redshift, which ignite a central starburst. The burst of star formation leads to a high central concentration of stars enshrouded in dust. In the centre of the resulting galaxy, a powerful quasar starts to clear out gas and dust from the central parts of the galaxy. The quasar quenches the star formation and a compact, quiescent galaxy is left behind.
This compact galaxy hereafter grows via encounters with minor companions: a mechanism called minor, dry merging. ``Dry'' here refers to the fact that the merging does not involve any significant gas to trigger new star formation. The scenario is summarized in Fig.~\ref{fig:galaxy_evolution}. Nevertheless, the proposed evolutionary scenario depicted above is highly debated. The exact mechanisms responsible for the size growth and for turning off star formation are still unknown.

Other mechanisms than merging have been proposed to explain the size growth, e.g., star formation at later times and quasar feedback; however, the most favoured mechanism has been minor merging. A cascade of merging events has been shown \citep{Oser2012} to effectively increase the half-light radius without changing the central density much (in agreement with observations).
Other studies, however, show that dry merging does not sufficiently describe the size evolution at all times \citep[e.g.,][]{Newman2012}.

At the same time, Newman et al. (2012) show that galaxies which formed at later times are larger on average. This is generally understood as a consequence of less gas-rich interactions at lower redshifts leading to less compact remnants.
This indicates that individual galaxies themselves do not need to evolve strongly in size. Instead, the quenching of larger galaxies at later times may increase the average size of the evolved population as a whole. In Chapter~\ref{K14}, we will study the effect of this ``dilution'' through a spectroscopically selected sample at redshift $z\approx 2$.

\subsection{Star-forming or Passively Evolving?}
\label{intro:UVJ}
\begin{figure}
	\centering
	\includegraphics[width=0.98\textwidth]{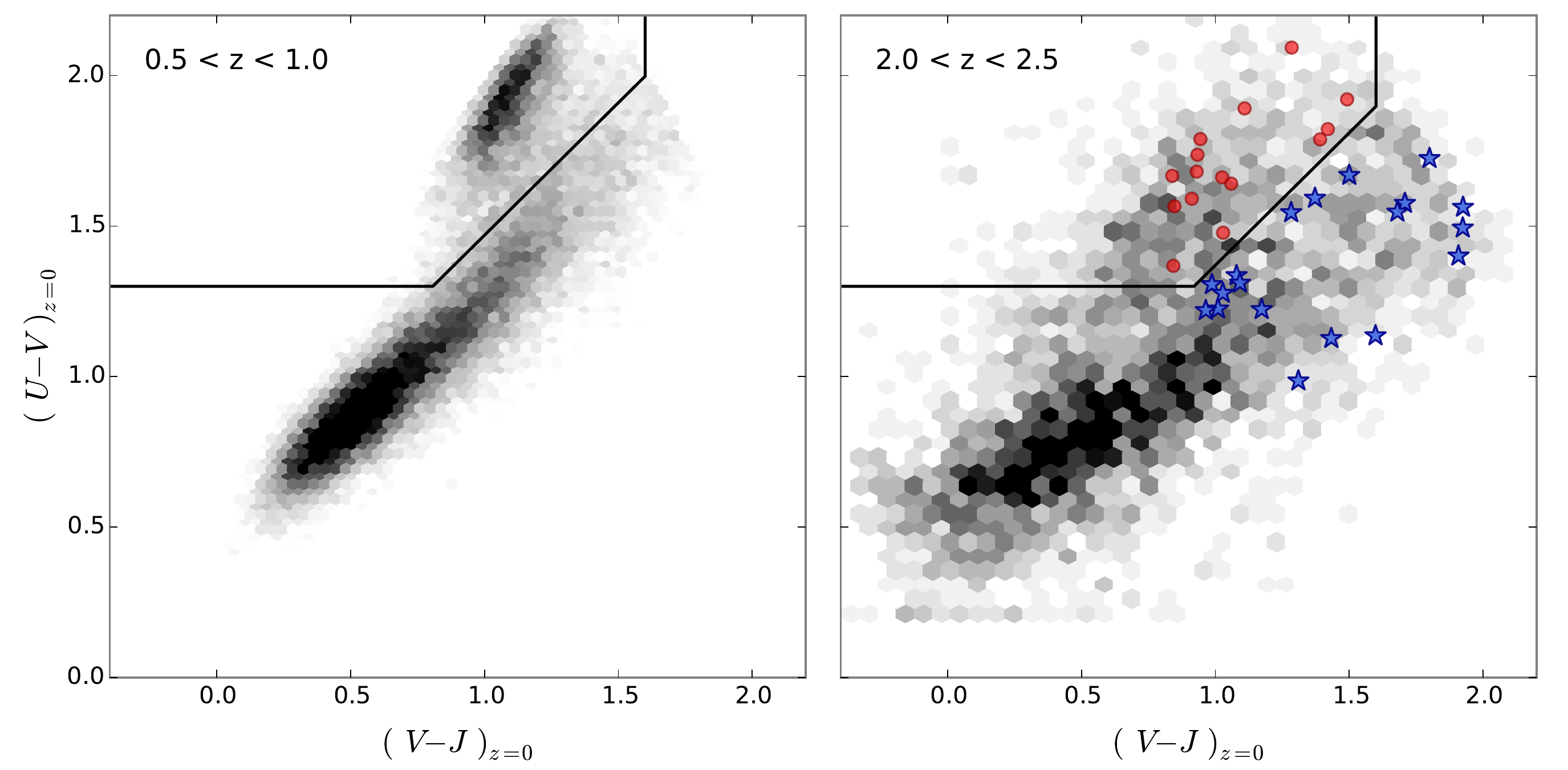}
	\caption{$UVJ$ rest-frame colour-colour diagram. The left panel
	shows data for intermediate redshifts $0.5<z<1.0$ and the right
	panel shows data for high redshifts $2.0<z<2.5$. The 2D-histograms
	show photometric data from \citet{Muzzin2013}. In each panel,
	the division between star-forming and quiescent from \citet{Williams2009}
	is shown by the black lines; quiescent galaxies reside in the
	upper left corner. In the right panel, the quiescent (red circles)
	and star-forming (blue stars) sub-samples from Chapter~\ref{K14} are shown.}
	\label{fig:UVJ}
\end{figure}

\begin{figure}
    \centering
	\includegraphics[width=0.55\textwidth]{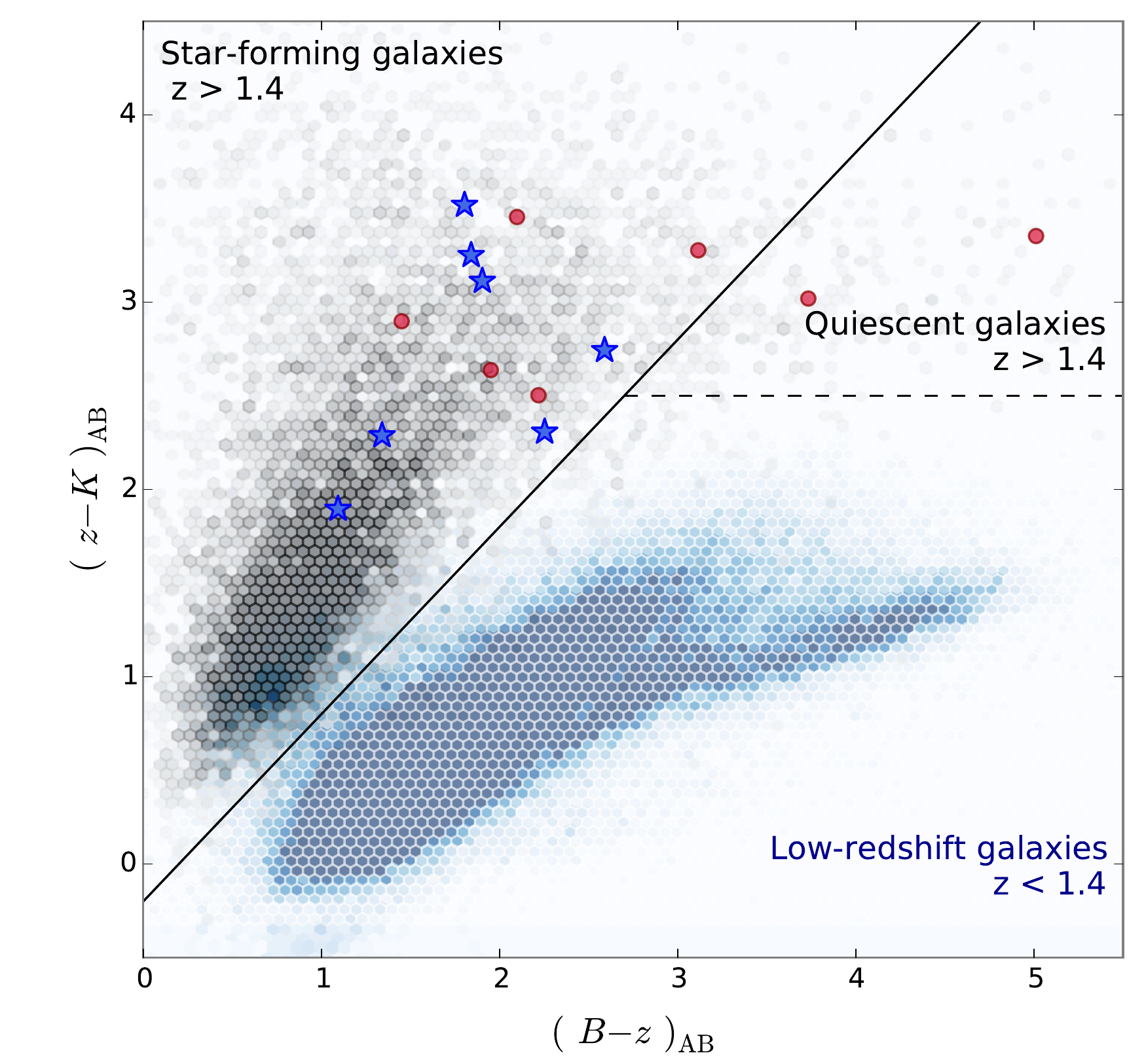}
	\caption{$B-z$ vs $z-K$ colour-colour diagram. Galaxies at low ($z<1$)
	and high ($1.9<z<2.5$) redshifts are shown by the blue and black
	2D-histograms, respectively. The photometric data are compiled from
	\citet{Muzzin2013}. The $BzK$ division between high and low redshift
	galaxies is shown by the black lines. Low-redshift galaxies reside
	in the lower right part of the diagram; high-redshift, quiescent galaxies
	reside in the upper right corner between the solid and dashed lines;
	and high-redshift, star-forming galaxies reside in the upper left corner.
	The quiescent (red circles) and star-forming (blue stars) sub-samples
	using the $UVJ$ method from Chapter~\ref{K14} are shown.
	The two sub-samples are not distinguishable using the $BzK$ method.}
	\label{fig:BzK}
\end{figure}

In order to study the quiescent galaxy population, one needs a way of separating the quiescent galaxies (QGs) from star-forming galaxies (SFGs). This is rather trivial at low redshift where the massive, red, elliptical galaxies are easily distinguished from blue, disc-dominated, star-forming galaxies. At redshift $z\sim2$, the distinction between the two classes becomes less clear as the galaxies are only barely resolved. Moreover, a large fraction of the star-forming galaxies is enshrouded in dust, making the star-forming galaxies appear red. For this reason, a simple one-colour criterion is not sufficient, and either an additional colour or magnitude must be included. The $UVJ$ selection is one such classification method out of many.
The classification is based on photometry in rest-frame $U-V$ and $V-J$ colours \citep{Labbe2005, Williams2009}, and the method is therefore model-dependent and relies on the assumed spectral shape in order to calculate the fluxes at the rest-frame wavelengths \citep[e.g.,][]{Brammer2008, Taylor2009}. The $UVJ$ method is illustrated in Figure~\ref{fig:UVJ}.
Another similar approach is the so-called $BzK$ method, which relies on the observed $B-z$ and $z-K$ colours: $BzK \equiv (z-K)_{\rm AB} - (B-z)_{\rm AB}$ \citep{Daddi2004a}. Since the $BzK$ method uses observed colours rather than rest-frame colours, it is not as such prone to errors introduced by the interpolation of rest-frame colours. However, the distinction between SFGs and QGs is more clearly defined over a large redshift span for the $UVJ$ method. For comparison, the $BzK$ diagram is shown in Figure~\ref{fig:BzK}. While the separation between low- and high-redshift galaxies is easily identified in the $BzK$ diagram, the distinction between star-forming and quiescent galaxies is less obvious.
In the $UVJ$ diagram, this separation into star-forming and quiescent galaxies is clearly identified for redshifts less than $z<2$ \citep{Williams2010}. Due to the enhanced sensitivity to star-formation rate in the $UVJ$ diagram, we have chosen to use the $UVJ$ method in Chapter~\ref{K14}. The photometric classification is bolstered by the recovered star formation rates from the full photometric and spectral modelling, see Chapter~\ref{K14}. Furthermore, the sources that have significant (i.e., more than 3 $\sigma$) detections at 24~$\mu$m are all classified as star-forming using the $UVJ$ method. 
Correspondingly, all the galaxies classified as quiescent from $UVJ$ method are either detected at less than 3 $\sigma$ or not detected at all. For reference, the 24~$\mu$m fluxes and the $UVJ$ classifications for the spectroscopic sample in Chapter~\ref{K14} are given in Table~\ref{tab:intro_mips24}.

\begin{table}
\centering
\caption{{\sc mips} 24~$\mu$m fluxes for the spectroscopic sample of Chapter~\ref{K14}.\label{tab:intro_mips24}}

\begin{threeparttable}
\begin{tabular}{p{2.5cm} c r}
\toprule
ID      & $F_{24\,\mu{\rm m}}$ & Classification$^{a}$ \\
        & {\small ($\mu$Jy)}   &    \\[0.4ex]
\midrule
118543  & $ 321.80 \pm  8.59$ &  SFG\\[1mm]
121761  & $  84.48 \pm  8.63$ &  SFG\\[1mm]
122398  & $<21.91\ (3\sigma)$ &   QG\\[1mm]
123235  & $  79.20 \pm  8.84$ &  SFG\\[1mm]
124666  & \ldots              &   QG\\[1mm]
126824  & $  65.21 \pm  8.77$ &  SFG\\[1mm]
127466  & $<24.65\ (3\sigma)$ &   QG\\[1mm]
127603  & $  43.61 \pm  8.30$ &  SFG\\[1mm]
128061  & \ldots              &   QG\\[1mm]
128093  & $ 149.55 \pm  7.79$ &  SFG\\[1mm]
129022  & \ldots              &   QG\\[1mm]
134068  & $ 489.63 \pm  8.47$ &  SFG\\[1mm]
135878  & \ldots              &   QG\\[1mm]
140122  & $<18.80\ (3\sigma)$ &   QG\\
\bottomrule
\end{tabular}

\begin{tablenotes}[para,flushleft]
    $^{a}$ based on $UVJ$ method.
\end{tablenotes}
\end{threeparttable}
\end{table}

\clearpage

\clearpage
\newpage
\clearpage

\leavevmode
\vspace{8cm}

\begin{center}
	{\sc\Large Note from the author}	
\end{center}

{\large
\noindent
In order to keep the arXiv submission short, the individual chapters are not given in full length. Instead the bibliographic information as well as a link to the refereed journal article, that each chapter is based on, is given together with the title, author list, and abstract.\\

\noindent
The journal articles for each chapter are also available on arXiv.org.
}

%% file: divisions/Krogager2012.tex
\chapter{Metal-rich damped Lyman-\boldmath{$\alpha$} absorbers as probes of faint galaxies}
\label{K12}

This chapter contains the following article:\\

{\renewcommand*{\thefootnote}{\large\fnsymbol{footnote}}

{\bf \Large ``On the sizes of \boldmath{$z\gtrsim2$} Damped
Ly\boldmath{$\alpha$} Absorbing Galaxies''}
\\
}

\noindent
{\small Published as a Letter in \href{http://adsabs.harvard.edu/cgi-bin/nph-abs_connect?fforward=http://dx.doi.org/10.1111/j.1745-3933.2012.01272.x}{\it Monthly Notices of the Royal Astronomical Society: Letters, vol. 424, pp. L1--L5, 2012.}
}\\

\noindent
Authors:\\
\indent J.-K. Krogager,
J. P. U. Fynbo,
P. M\o ller,
C. Ledoux,
P. Noterdaeme,\\
\indent L. Christensen,
B. Milvang-Jensen,
\& M. Sparre\\[1cm]

Recently, the number of detected galaxy counterparts of $z \gtrsim 2$ damped
Lyman $\alpha$ absorbers in QSO spectra has increased substantially so we
today have a sample of 10 detections. M\o ller et al.\ in the year 2004 made the
prediction, based on a hint of a luminosity--metallicity relation for DLAs, that
\ion{H}{i} size should increase with increasing metallicity. In this Letter we
investigate the distribution of impact parameter and metallicity that would result
from the correlation between galaxy size and metallicity. We compare our observations
with simulated data sets given the relation of size and metallicity.
The observed sample presented here supports the metallicity--size prediction:
The present sample of DLA galaxies is consistent with the model distribution.
Our data also show a strong relation between impact parameter and
column density of \ion{H}{i}.
We furthermore compare the observations with several numerical simulations and demonstrate that the observations support a scenario where the relation between size and metallicity is driven by feedback mechanisms controlling the star-formation efficiency and outflow of enriched gas.\\

\begin{figure}
    \centering
	\includegraphics[width=1\textwidth]{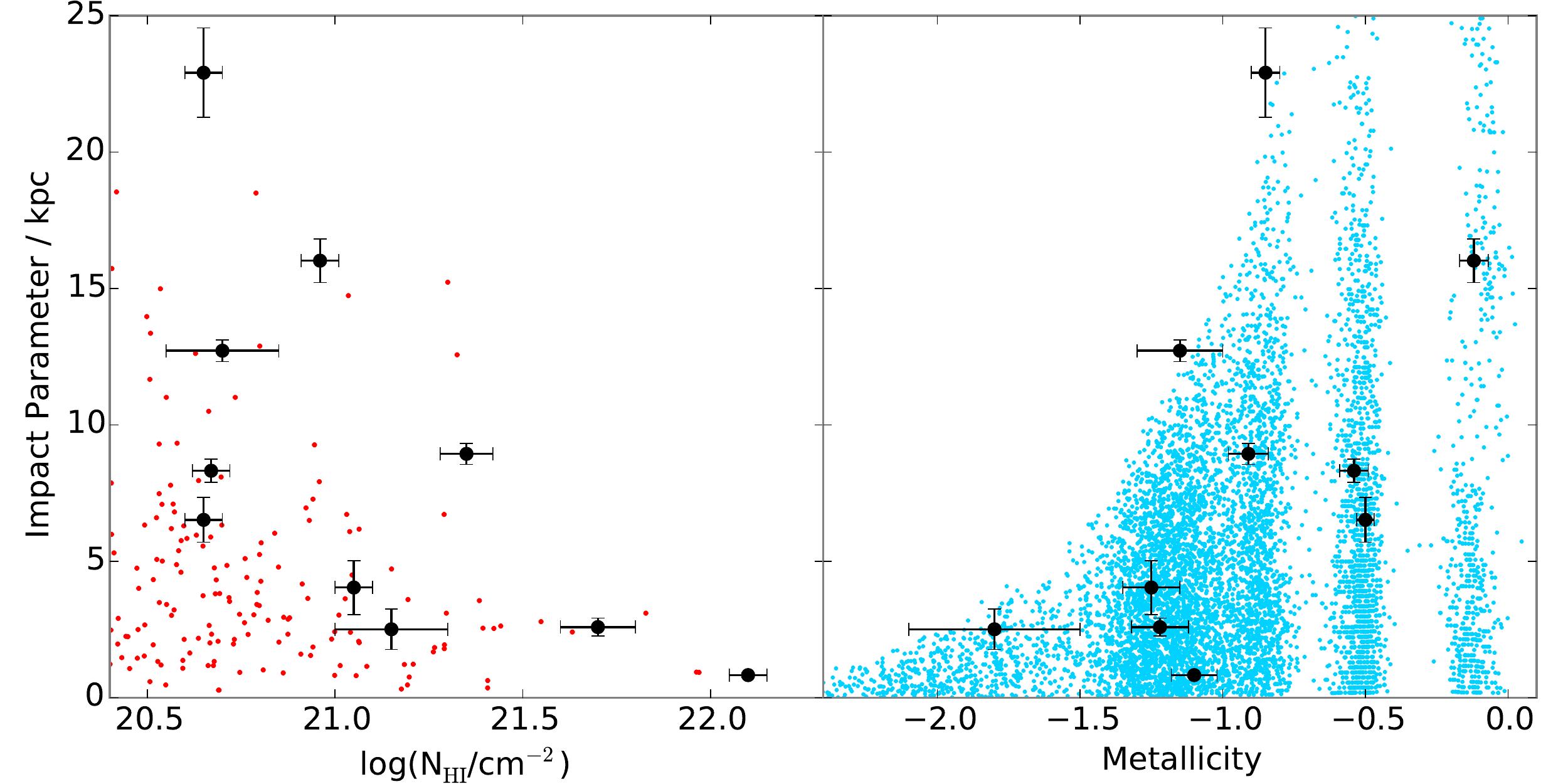}
	\caption{Impact parameters plotted against \ion{H}{i} column density ({\it left})
	and metallicity ({\it right}) for $z\gtrsim2$ DLAs with securely identified galaxy
	counterparts (see table~\ref{DLAtable}). The observed impact parameters have been 
	converted to kpc at redshift $z=3$ instead of arcsec for easier comparison.
	The blue points in the right panel show the simulated distribution of
	impact parameters as a function of metallicity for DLA galaxies at $z=3$
	from the model in \citet{Fynbo2008}. The red points in the left panel
	show model points from \citet{Pontzen2008}.\\
	{\it -- Updated version of Figure 3 in the published article.}
	\label{fig:bZplot}}
\end{figure}

\begin{figure}
    \centering
	\includegraphics[trim=0mm 0mm 0mm 0mm,clip,width=0.95\textwidth]{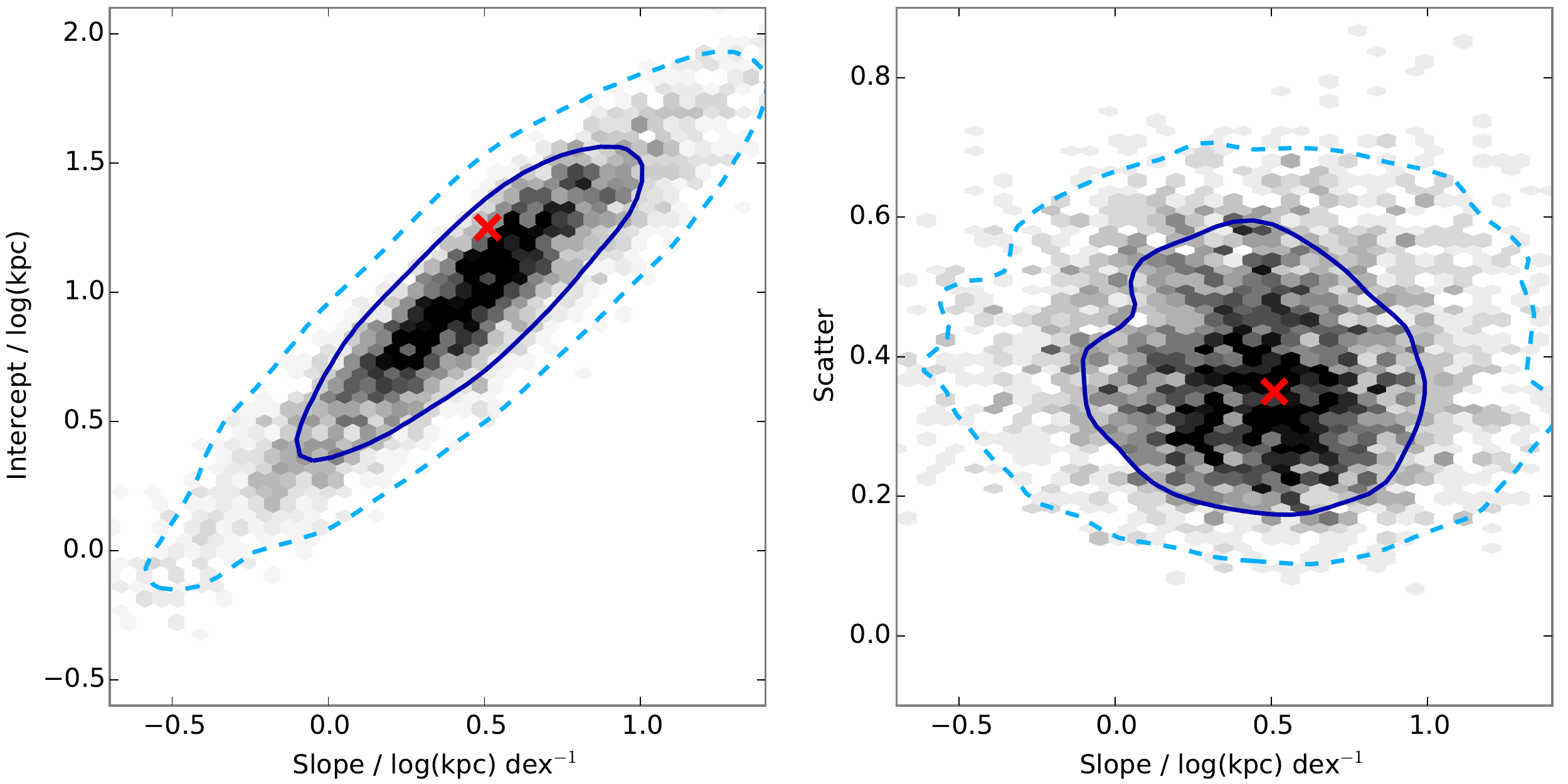}
	\caption{\small Distribution of slope vs. intercept ({\it left}) and
	slope vs. scatter ({\it right}) from the 4\,000 simulated
	data sets (grey 2D-histogram). The solid and dashed blue lines represent the 1$\sigma$
	and 2$\sigma$ confidence contours, respectively. The red cross shows
	the result from our fit to the data.\\
	{\it -- Updated version of Figure 4 in the published article.}
	\label{fig:hist}}
\end{figure}

\noindent
{{\sc Note:} The metallicities and column densities of \ion{H}{i}
in Table~\ref{DLAtable} have been updated relative to the published version
to use the measurements from \citet{Ledoux2006}, where available, since these are measured
more homogeneously.
The values have also been updated in Figures~\ref{fig:bZplot} and \ref{fig:hist}.
The results of the analysis and our conclusions remain unaltered.\\
The affected Table and Figures are shown in the following pages.}

\newpage

\begin{table}
\centering
\caption{$z\gtrsim2$ DLAs with identified galaxy counterparts used in this study
\label{DLAtable}}

\begin{threeparttable}
\begin{tabular}{p{3.5cm} c c c c}
\toprule
QSO   & $z_{\rm abs}$ & $\log(N_{\rm H \textsc{i}}/{\rm cm}^{-2})$ & [M/H] &  $b$     \\
   	  &               &                                            &       & (arcsec) \\[0.4ex]
\midrule
Q2206$-$19$^{(1,12)}$	 & 1.92 & 20.67 $\pm$ 0.05 & $-$0.54 $\pm$ 0.05\,$^{\mathrm{Zn}}$ & 0.99 $\pm$ 0.05 \\
PKS0458$-$02$^{(1,5,12)}$& 2.04 & 21.70 $\pm$ 0.10 & $-$1.22 $\pm$ 0.10\,$^{\mathrm{Zn}}$ & 0.31 $\pm$ 0.04 \\
Q1135$-$0010$^{(7)}$     & 2.21 & 22.10 $\pm$ 0.05 & $-$1.10 $\pm$ 0.08\,$^{\mathrm{Zn}}$ & 0.10 $\pm$ 0.01 \\
Q0338$-$0005$^{(5,8)}$   & 2.22 & 21.05 $\pm$ 0.05 & $-$1.25 $\pm$ 0.10\,$^{\mathrm{Si}}$ & 0.49 $\pm$ 0.12 \\
Q2243$-$60$^{(4,12)}$	 & 2.33 & 20.65 $\pm$ 0.05 & $-$0.85 $\pm$ 0.05\,$^{\mathrm{Zn}}$ & 2.80 $\pm$ 0.20 \\
Q2222$-$0946$^{(2,12)}$ 	 & 2.35 & 20.65 $\pm$ 0.05 & $-$0.50 $\pm$ 0.03\,$^{\mathrm{Zn}}$ &  0.8 $\pm$ 0.1  \\
Q0918$+$1636$^{(3,5)}$	 & 2.58 & 20.96 $\pm$ 0.05 & $-$0.12 $\pm$ 0.05\,$^{\mathrm{Zn}}$ &  2.0 $\pm$ 0.1  \\
Q0139$-$0824$^{(6,9)}$   & 2.67 & 20.70 $\pm$ 0.15 & $-$1.15 $\pm$ 0.15\,$^{\mathrm{Si}}$ & 1.60 $\pm$ 0.05 \\
PKS0528$-$250$^{(1,12)}$ & 2.81 & 21.35 $\pm$ 0.07 & $-$0.91 $\pm$ 0.07\,$^{\mathrm{Zn}}$ & 1.14 $\pm$ 0.05 \\
Q0953$+$47$^{(10,11)}$	 & 3.40 & 21.15 $\pm$ 0.15 & $-$1.80 $\pm$ 0.30\,$^{\mathrm{Si}}$ & 0.34 $\pm$ 0.10 \\
\bottomrule
\addlinespace[0.6ex]
\end{tabular}

\begin{tablenotes}[para,flushleft]
\small
{\bf References:} (1) \citet{Moller2004}; (2) \citet{Fynbo2010}; (3) \citet{Fynbo2011};\\
(4) \citet{Bouche2012a}; (5) this work; (6) Christensen et al.\ in prep.;\\
(7) \citet{Noterdaeme2012a}; (8) Ledoux, priv. comm.; (9) \citet{Wolfe2008};\\
(10) A. Bunker (priv. communication); (11) \citet{Prochaska2003a}; (12) \citet{Ledoux2006}.\\
{\it -- Updated version of Table 1 in the published article.}
\end{tablenotes}
\end{threeparttable}
\end{table}

\clearpage

%% file: divisions/Krogager2013.tex
\chapter{Emission from a metal-rich DLA: Evidence for enriched galactic outflows}
\label{K13}

This chapter contains the following article:\\

\hspace{-0.24cm}{\bf{\Large ``Comprehensive Study of a \boldmath{$z=2.35$} DLA Galaxy:}\\
\indent {\large Mass, Metallicity, Age, Morphology and SFR from HST and VLT}''}\\

\noindent
Published in \href{http://adsabs.harvard.edu/cgi-bin/nph-abs_connect?fforward=http://dx.doi.org/10.1093/mnras/stt955}{\it Monthly Notices of the Royal Astronomical Society, vol. 433, pp. 3091-3102, 2013.}\\

\noindent
Authors:\\
\indent J.-K. Krogager,
J. P. U. Fynbo,
C. Ledoux,
L. Christensen,
A. Gallazzi,\\
\indent P. Laursen,
P. M\o ller,
P. Noterdaeme,
C. P\'eroux,
M. Pettini,
\& M. Vestergaard\\

We present a detailed study of the emission from a $z=2.35$ galaxy that
causes damped Lyman $\alpha$ absorption in the spectrum of the background
quasar, SDSS J\,2222$-$0946. We present the results of extensive analyses
of the stellar continuum covering the rest-frame UV--optical regime based
on broad-band {\it Hubble Space Telescope (HST)} imaging, and of
spectroscopy from VLT/X-shooter of the strong emission lines: Ly$\alpha$,
[\ion{O}{ii}], [\ion{O}{iii}], [\ion{N}{ii}], H$\alpha$, and H$\beta$.
We compare the metallicity from the absorption lines in the quasar spectrum with the oxygen abundance inferred from the strong-line methods ($R_{23}$
and N2). The two emission-line methods yield consistent results:
[O/H] = $-0.30\pm0.13$. Based on the absorption lines in the quasar
spectrum a metallicity of $-0.49\pm0.05$ is inferred at an impact parameter
of 6.3 kpc from the centre of the galaxy with a column density of hydrogen
of $\log(N_{\rm H \textsc{i}} / {\rm cm}^{-2})=20.65\pm0.05$.
The star formation rate of the galaxy from the UV continuum and the
H$\alpha$ line can be reconciled assuming an amount of reddening of
$E(B-V) = 0.06\pm0.01$, giving an inferred star formation rate of
$13\pm1~\mathrm{M}_{\odot}$~yr$^{-1}$ (Chabrier IMF).
From the {\it HST} imaging, the galaxy associated with the absorption
is found to be a compact ($r_e$=1.12~kpc) object with a disc-like,
elongated (axis ratio 0.17) structure indicating that the galaxy is seen
close to edge-on. Moreover, the absorbing gas is located almost
perpendicularly above the disc of the galaxy suggesting that the gas
causing the absorption is not co-rotating with the disc. We investigate the
stellar and dynamical masses from SED-fitting and emission-line widths,
respectively, and find consistent results of $2 \times10^9$~M$_{\odot}$.
We suggest that the galaxy is a young {\it proto}-disc with evidence for
a galactic outflow of enriched gas. This galaxy hints at how star-forming
galaxies may be linked to the elusive population of damped Ly$\alpha$
absorbers.

\newpage

%% file: divisions/redQSOs.tex
\chapter{Identifying Red Quasars}
\label{redQSOs}

This chapter contains the following article:\\

\hspace{-0.24cm}{\Large\bf ``Optical/Near-Infrared Selection of Red Quasi-Stellar Objects:\\
\indent {\large Evidence for Steep Extinction Curves toward Galactic Centers?}''}\\

\noindent
Published in \href{http://adsabs.harvard.edu/cgi-bin/nph-abs_connect?fforward=http://dx.doi.org/10.1088/0067-0049/204/1/6}{\it The Astrophysical Journal Supplement Series, vol. 204, id. 6, 14 pp., 2013.}\\

\noindent
Authors:\\
\indent J.~P.~U.~Fynbo, J.-K. Krogager, B. Venemans, P. Noterdaeme, M. Vestergaard,\\
\indent P. M{\o}ller, C. Ledoux, \& S. Geier\\

We present the results of a search for red QSOs using a selection based on
optical imaging from SDSS and near-infrared imaging from UKIDSS. Our main goal
with the selection is to search for QSOs reddened by foreground dusty absorber
galaxies. For a sample of 58 candidates (including 20 objects fulfilling our
selection criteria that already have spectra in the SDSS) 46 (79\%) are
confirmed to be QSOs. The QSOs are predominantly dust-reddened except for a handful
at redshifts $z\gtrsim3.5$. However, the dust is most likely located in the
QSO host galaxies (and for two the reddening is primarily caused by Galactic
dust) rather than in intervening absorbers. More than half of the QSOs show
evidence of associated absorption (BAL absorption). Four (7\%) of the candidates
turned out to be late-type stars, and another four (7\%) are compact galaxies. We could not identify the remaining four objects. 
In terms of their optical spectra, these QSOs are similar to the QSOs selected in the FIRST-2MASS Red Quasar Survey except they are on average  fainter, more distant, and only two are detected in the FIRST survey. As per the usual procedure, we estimate the amount of extinction using the SDSS QSO template reddened by SMC-like dust.
It is possible to get a good match to the observed (restframe ultraviolet) spectra, but it is not possible to match the observed near-IR photometry from UKIDSS for nearly all the reddened QSOs.
The most likely reasons are that the SDSS QSO template is too red at optical wavelengths due to contaminating host galaxy light and that the assumed SMC extinction curve is too shallow. Three of the compact galaxies display old stellar populations with ages of several Gyr and masses of about 10$^{10}$ M$_{\odot}$ (based on spectral energy distribution modelling).
The inferred stellar densities in these galaxies exceed 10$^{10}$~M$_{\odot}$ kpc$^{-2}$, which is among the highest measured for early type galaxies. Our survey has demonstrated that selection of QSOs based on near-IR photometry is an efficient way to select QSOs, including reddened QSOs, with only small contamination from late-type stars and compact galaxies.

\newpage

%% file: divisions/HAQ.tex
\newcommand{\total}{$159$}

\chapter{Dust in intervening absorption systems toward red quasars}
\label{HAQ}

This chapter contains the following article:\\

\hspace{-0.24cm}{\bf\Large ``The High \boldmath{$A_V$} Quasar Survey: Reddened quasi-stellar objects\\
\indent selected from optical/near-infrared photometry -- II''}\\

\noindent
Published in \href{http://adsabs.harvard.edu/cgi-bin/nph-abs_connect?fforward=http://dx.doi.org/10.1088/0067-0049/217/1/5}{\it The Astrophysical Journal Supplement Series, vol. 217, id. 5, 26 pp, 2015.}\\

\noindent
Authors:\\
\indent J.-K. Krogager,
S. Geier,
J. P. U. Fynbo,
B. P. Venemans,
C. Ledoux,
P. M\o ller,\\
\indent P. Noterdaeme,
M. Vestergaard,
T. Kangas,
T. Pursimo,
F. G. Saturni,
\& O. Smirnova\\[1cm]

Quasi-stellar objects (QSOs) whose spectral energy distributions (SEDs) are reddened by dust either in their host galaxies or in intervening absorber galaxies are to a large degree missed by optical colour
selection criteria like the one used by the Sloan Digital Sky Survey (SDSS). To overcome this bias against red QSOs, we employ a combined optical and near-infrared colour selection.
In this paper, we present a spectroscopic follow-up campaign of a sample of red
candidate QSOs which were selected from the SDSS and the UKIRT Infrared Deep Sky Survey (UKIDSS).
The spectroscopic data and SDSS/UKIDSS photometry are
supplemented by mid-infrared photometry from the Wide-field Infrared Survey Explorer. In
our sample of 159 candidates, 154 (97\%) are confirmed to be QSOs.
We use a statistical algorithm to identify sightlines with plausible intervening absorption systems and identify nine such cases assuming dust in the absorber similar to Large Magellanic Cloud sightlines.
We find absorption systems toward 30 QSOs, 2 of which are consistent with the best-fit absorber redshift from the statistical modelling.
Furthermore, we observe a broad range in SED properties of the QSOs as probed by the rest-frame 2~$\mu$m flux. We find QSOs with a strong excess as well as QSOs with a large deficit at rest-frame 2~$\mu$m relative to a QSO template. Potential solutions to these discrepancies are discussed. Overall, our study demonstrates the
high efficiency of the optical/near-infrared selection of red QSOs.

\newpage

%% file: divisions/Krogager2015.tex
\chapter{The tip of the dusty iceberg}
\label{K15}

This chapter contains the following article:\\

\hspace{-0.23cm}{\bf \Large ``A Quasar reddened by a sub-parsec sized, metal-rich and\\
\indent dusty cloud in a damped Lyman-\boldmath{$\alpha$} absorber at $z=2.13$''}\\

\noindent
Published in \href{http://adsabs.harvard.edu/cgi-bin/nph-abs_connect?fforward=http://dx.doi.org/10.1093/mnras/stv2346}{\it Monthly Notices of the Royal Astronomical Society, vol. 455, pp. 2698-2711, 2016}.\\

\noindent
Authors:\\
\indent J.-K. Krogager,
J. P. U. Fynbo,
P. Noterdaeme,
T. Zafar,
P. M\o ller,\\
\indent C. Ledoux,
T. Kr\"uhler,
\& A. Stockton\\[1cm]

We present a detailed analysis of a red quasar at $z=2.32$ with an intervening damped Lyman $\alpha$ absorber (DLA) at $z=2.13$. Using high-quality data from the X-shooter spectrograph at ESO Very Large Telescope, we find that the absorber has a metallicity consistent with solar. We observe strong \ion{C}{i} and H$_2$ absorption indicating a cold, dense absorbing medium. Partial coverage effects are observed in the \ion{C}{i} lines, from which we infer a covering fraction of $27\pm6$ per cent and a physical diameter of the cloud of 0.1~pc. From the covering fraction and size, we estimate the size of the background quasar’s broad line region. We search for emission from the DLA counterpart in optical and near-infrared imaging. No emission is observed in the optical data. However, we see tentative evidence for a counterpart in the $H$- and $K′$-band images. The DLA shows high depletion (as probed by $[{\rm Fe/Zn}]=-1.22$) indicating that significant amounts of dust must be present in the DLA. By fitting the spectrum with various dust reddened quasar templates, we find a best-fitting amount of dust in the DLA of ${\rm A(V)_{DLA}}=0.28\pm0.01|_{\rm stat}\ \pm0.07|_{\rm sys}$. We conclude that dust in the DLA is causing the colours of this intrinsically very luminous background quasar to appear much redder than average quasars, thereby not fulfilling the criteria for quasar identification in the Sloan Digital Sky Survey. Such chemically enriched and dusty absorbers are thus under-represented in current samples of DLAs.

\newpage

%% file: divisions/Krogager2014.tex
\chapter{The size evolution of massive, evolved galaxies from redshift 2 to 0}
\label{K14}

This chapter contains the following article:\\

\noindent\hspace{-0.24cm}{\bf\Large ``A spectroscopic sample of massive, quiescent \boldmath{$z\sim2$} galaxies:\\
Implications for the evolution of the mass--size relation''}\\

\noindent
Published in \href{http://adsabs.harvard.edu/cgi-bin/nph-abs_connect?fforward=http://dx.doi.org/10.1088/0004-637X/797/1/17}{\it The Astrophysical Journal, vol. 797, id. 17, 14 pp., 2014.}\\

\noindent
Authors:\\
\indent J.-K. Krogager,
A. W. Zirm,
S. Toft,
A. Man,
\& G. Brammer\\[4mm]

We present deep, near-infrared {\it Hubble Space Telescope}/Wide Field Camera 3 grism spectroscopy and imaging for
a sample of 14 galaxies at $z\approx2$ selected from a mass-complete photometric
catalog in the COSMOS field.
By combining the grism observations with photometry in 30 bands, we derive
accurate constraints on their redshifts, stellar masses, ages, dust
extinction and formation redshifts. 
We show that the slope and scatter of the $z\sim2$ mass--size
relation of quiescent galaxies is consistent with the local relation, and confirm previous
findings that the sizes for a given mass are smaller by a factor of two to three. 
Finally, we show that the observed evolution of the mass--size relation of
quiescent galaxies between $z=2$ and $0$ can be explained by 
quenching of increasingly larger star-forming galaxies at a rate
dictated by the increase in the number density of quiescent
galaxies with decreasing redshift. However, we find that the scatter in the mass--size
relation should increase in the quenching-driven scenario in contrast to what is seen
in the data. This suggests that merging is not needed to explain the
evolution of the median mass--size relation of massive galaxies, but may
still be required to tighten its scatter, and explain the size growth
of individual $z=2$ quiescent galaxies.

\newpage

%% file: divisions/summary/conclusion.tex
\chapter{Conclusions and Outlook}
\label{conclusion}

In Chapter~\ref{K12}, we studied the elusive nature of galaxies causing damped Ly$\alpha$ absorbers. In order to enhance the chances of detecting the faint glow from the DLA galaxies, we targeted the metal-rich absorbers only as these had been hypothesized to have more luminous counterparts. Indeed, this approach led to 4 detections out of a sample of 9 DLAs. We found that the relation between metallicity and impact parameter was consistent with DLAs being drawn from the low-luminosity population of star-forming galaxies (Lyman-break galaxies, LBGs).
In Chapter~\ref{K13}, one of these metal-rich absorbers was studied in high detail using state-of-the-art instrumentation both from ground and space. This revealed a small, actively star-forming galaxy associated with the absorption. By modelling the \lya\ emission we found that outflows must be important to explain the line profile. We therefore concluded that the DLA was caused by enriched material blown out from the star-forming galaxy. Furthermore, we were able to measure the stellar mass of the galaxy: $M_{\star}=2\times10^{9}~{\rm M_{\odot}}$. This was the first direct measurement of the stellar mass of a high-redshift DLA galaxy.

Although the DLAs presented in Chapter~\ref{K12} make up a heterogeneous and biased sub-sample of the general DLA population, the increase in sample size allowed us to start probing the underlying nature of DLAs in direct emission \citep[see][]{Christensen2014}.
So far, the thesis has only presented the detections of emission counterparts in the sample. The full analysis including the non-detections will be presented in a future paper. This will provide a more complete census of metal-rich DLAs, which can be more easily compared to other studies of the DLA population.

Recently, \citet{Fumagalli2014, Fumagalli2015} presented a sample of 32 DLAs with which they study {\it in-situ} star formation in the far-UV by imaging below the Lyman-limit caused by unrelated absorbers at higher redshifts. The authors find that the overall star formation rate associated directly with DLAs is very low ${\rm SFR}<2~{\rm M_{\odot}~yr^{-1}}$; however, these rates do not take into account the possible effects of dust and intergalactic absorption. These results are consistent with a scenario in which DLAs trace the neutral gas around low-luminosity star-forming galaxies. The increasing amount of evidence for an underlying mass--metallicity relation for DLAs \citep{Moller2013, Neeleman2013, Christensen2014} further corroborates these claims: the average DLA will be associated with galaxies of low mass and low luminosity, whereas the more metal-rich DLAs will belong to increasingly larger haloes hosting more massive and luminous galaxies. This picture is furthermore supported by recent cross-correlation analyses and simulations of DLAs within the $\Lambda$CDM paradigm \citep[e.g.,][]{FontRibera2012, Barnes2014, Bird2014, Bird2015}. Although the picture is not fully painted, the overall frame is present and most of the main observables are matched well.\\

\begin{figure}
    \centering
    \includegraphics[width=0.65\textwidth]{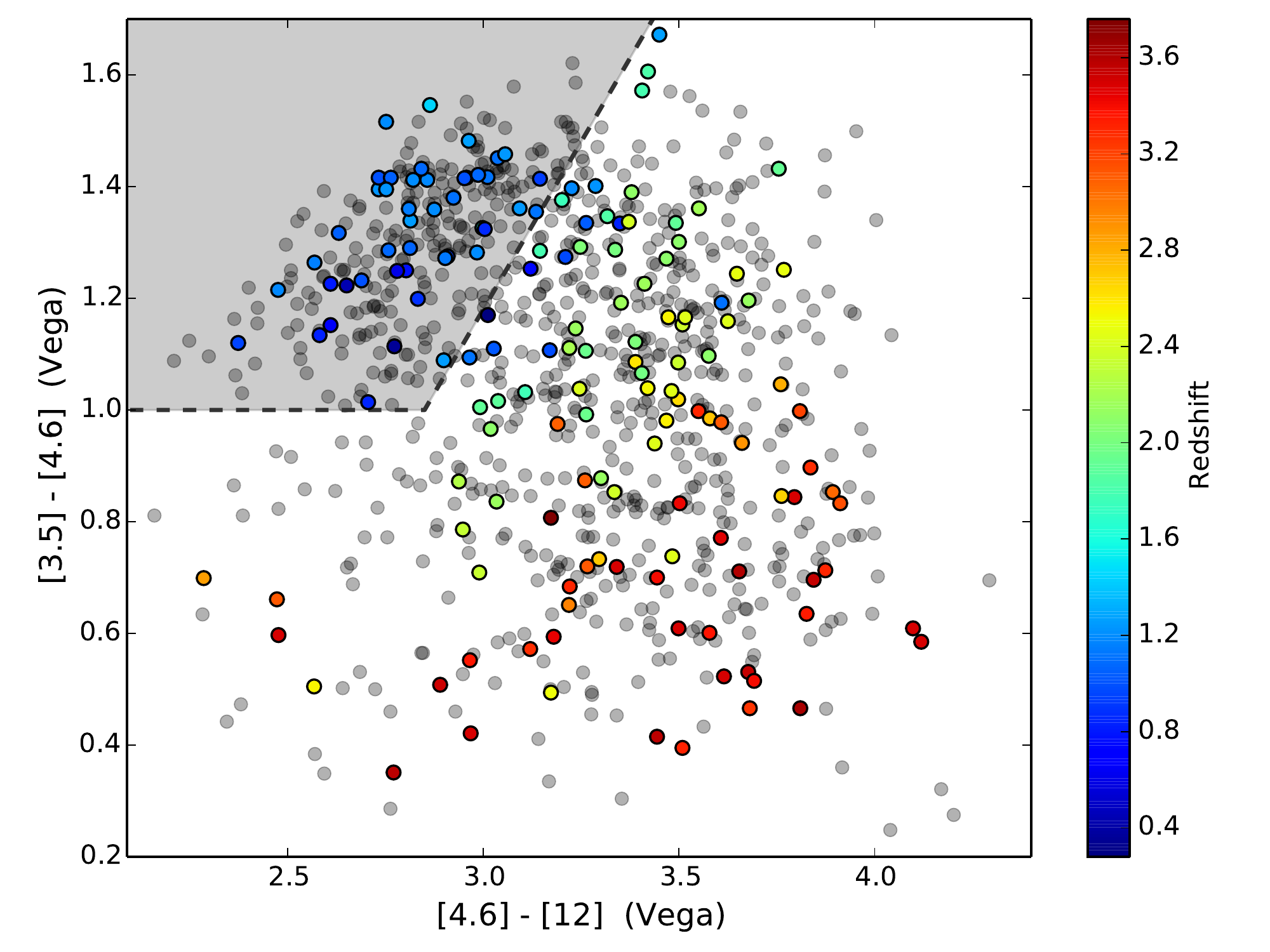}
    \caption{Colour-colour plot from WISE photometry. The full HAQ sample is shown as grey points
    and the spectroscopically confirmed quasars are shown by coloured points, where the colour
    indicates the quasar redshift. The grey shaded area in the upper left corner is the exclusion
    region for the eHAQ survey introduced to remove low-redshift interlopers.
    \label{fig:WISE_colors}}
\end{figure}

One aspect of understanding the nature of DLAs, which remains unanswered, is the question of dust in DLAs and whether our samples are systematically missing the most chemically enriched systems. This led us to the red quasar surveys presented in Chapters~\ref{redQSOs} and \ref{HAQ}. We found that most of the targeted quasars were not reddened by intervening absorbers; however, in Chapter~\ref{HAQ}, we did find a few quasars with foreground absorbers potentially harbouring dust. One of the quasars from the HAQ survey (QSO J\,2225+0527) was studied in detail in Chapter~\ref{K15}. We here found that the dust was indeed located in the absorber and not in the quasar. Furthermore, the background quasar was not identified in the SDSS database due to the strong reddening from the foreground dust. This shows that dusty DLAs are missed by optical quasar selection criteria. The impact of this bias will be studied in future projects where we will compare the different selection functions of SDSS and our surveys to the resulting quasar+DLA population.
We have initiated an extension of the HAQ survey (the eHAQ survey) to target quasars at higher redshifts and with larger amounts of reddening by selecting quasars from their mid-infrared properties. We use the near- and mid-infrared photometry from UKIDSS and WISE to separate stars and galaxies from quasars. The WISE photometry furthermore allows us to discard low-redshift quasars, see Figure~\ref{fig:WISE_colors}.
A similar approach was recently published on arXiv.org (Richards et al. 2015, arXiv:1507.07788, Jul 28$^{\rm th}$, {\it accepted for publication in ApJS}).

\begin{figure}
    \centering
    \includegraphics[width=0.7\textwidth]{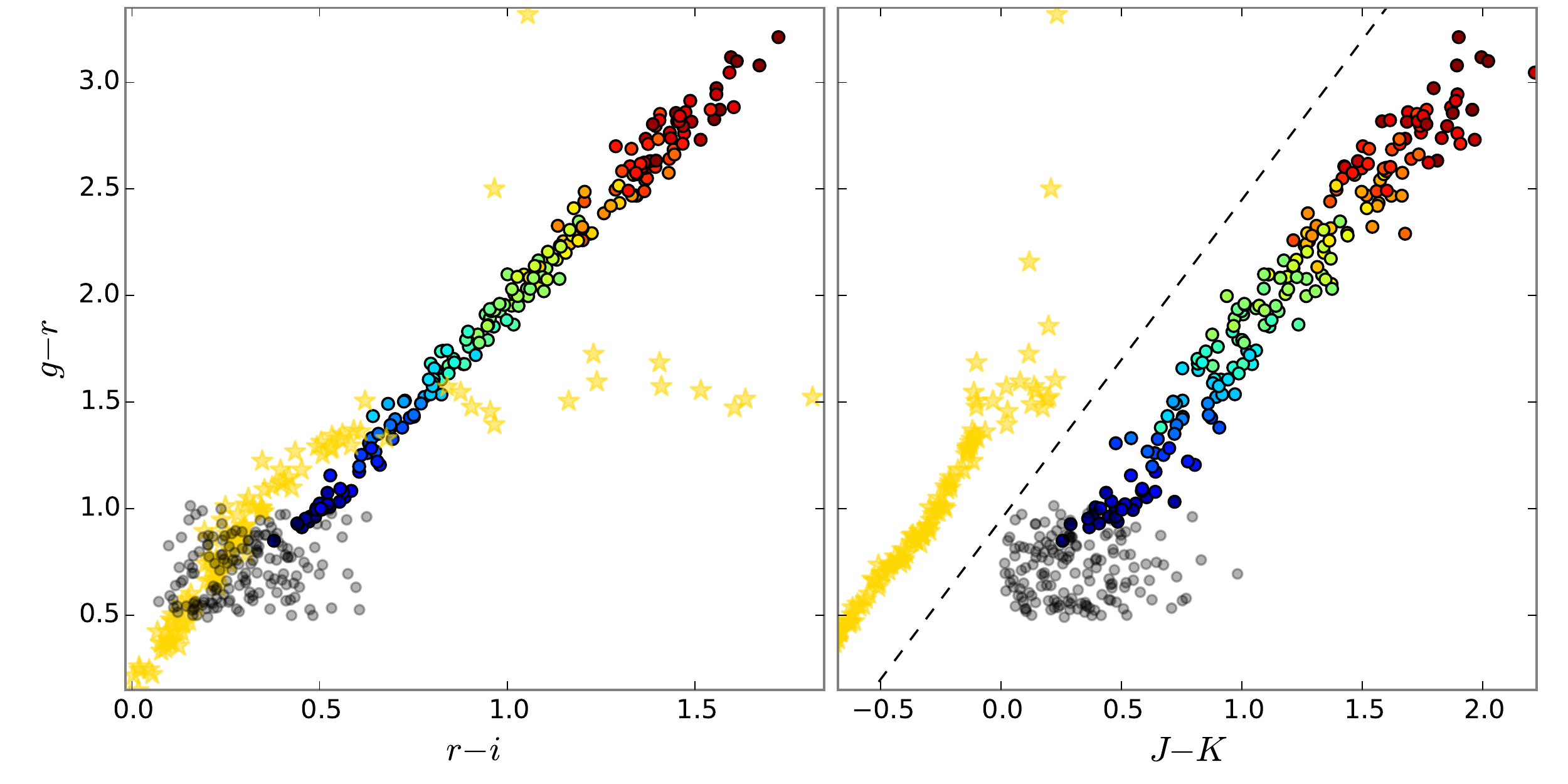}
    \includegraphics[width=0.7\textwidth]{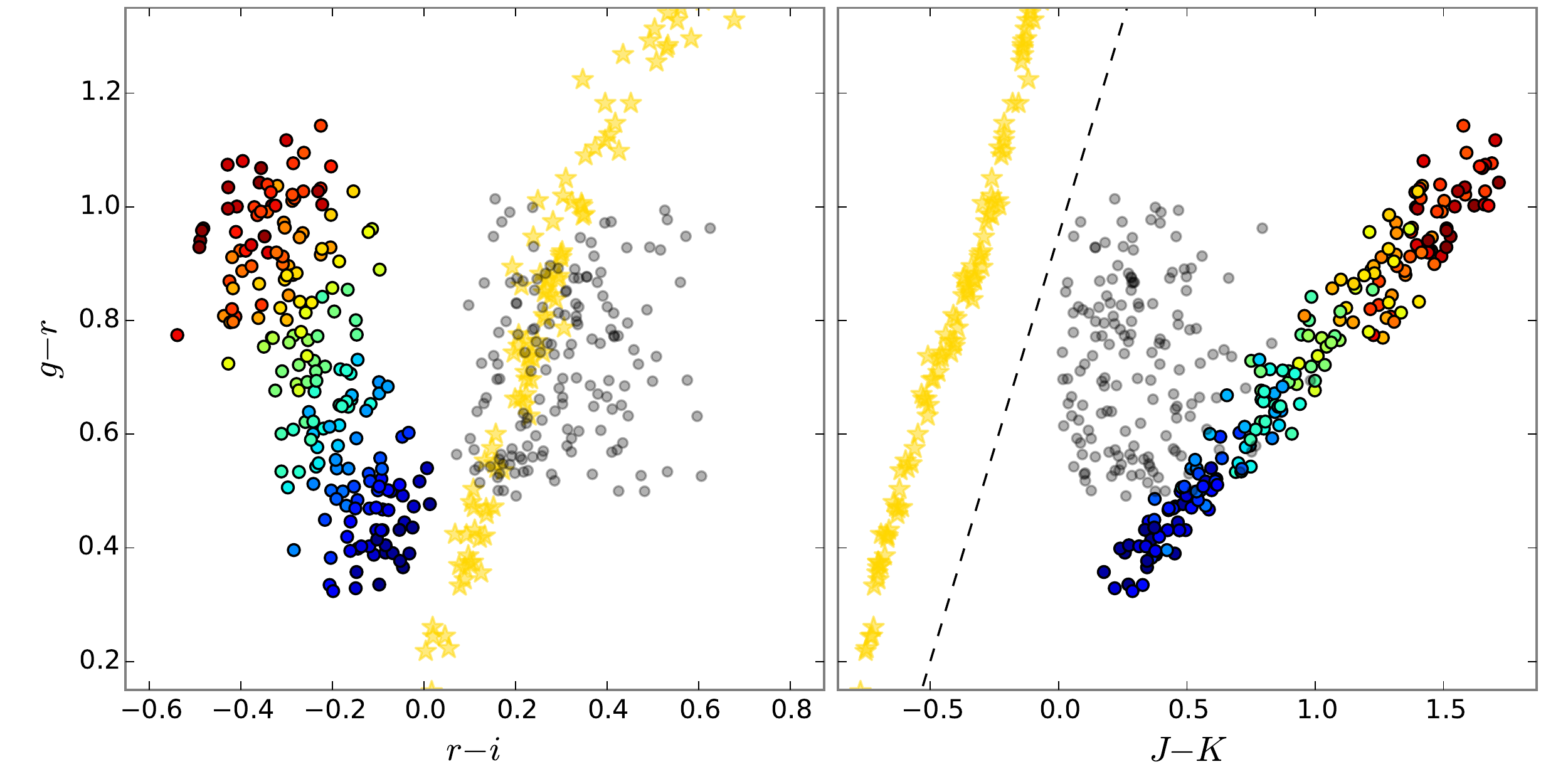}
    \caption{Colour-colour diagrams for simulated quasars with intervening DLAs.
    The coloured points indicate quasars at $z\approx3$ where the colour corresponds to
    the amount of reddening $A(V)$ (blue for $A(V)=0.5$ and red for $A(V)=2.0$).
    The grey points show the quasars from the HAQ survey and the yellow stars indicates
    the stellar locus taken from \citet{Hewett2006}.
    The top row shows colours assuming the SMC extinction curve in the DLA,
    and the bottom row assumes the LMC extinction curve in the DLA.
    The combination of optical ($g-r$) and near-infrared ($J-K$) colours clearly separates
    quasars from stars (indicated by the dashed line).
    \label{fig:eHAQ_colors}}
\end{figure}

A key goal for the eHAQ survey is to identify {\it redder} quasars than the HAQ survey and at higher redshifts. This will allow us to constrain the incidence of dusty DLAs to a larger degree than what was possible in the HAQ survey. In order to look for more dusty absorbers we will need to extend the survey to more red quasars in terms of their $g-r$ colour. This was originally introduced to have sufficient flux in the blue part of the spectra to look for \lya\ absorption.
We have simulated the colours of quasars at $z\approx3$ with foreground DLAs at $z_{\rm DLA}=2.5$ causing various amounts of reddening, in order to have an idea of the expected behaviour in colour-colour space. These simulated colours are shown in Figure~\ref{fig:eHAQ_colors}.


\begin{figure}
    \centering
    \includegraphics[width=\textwidth]{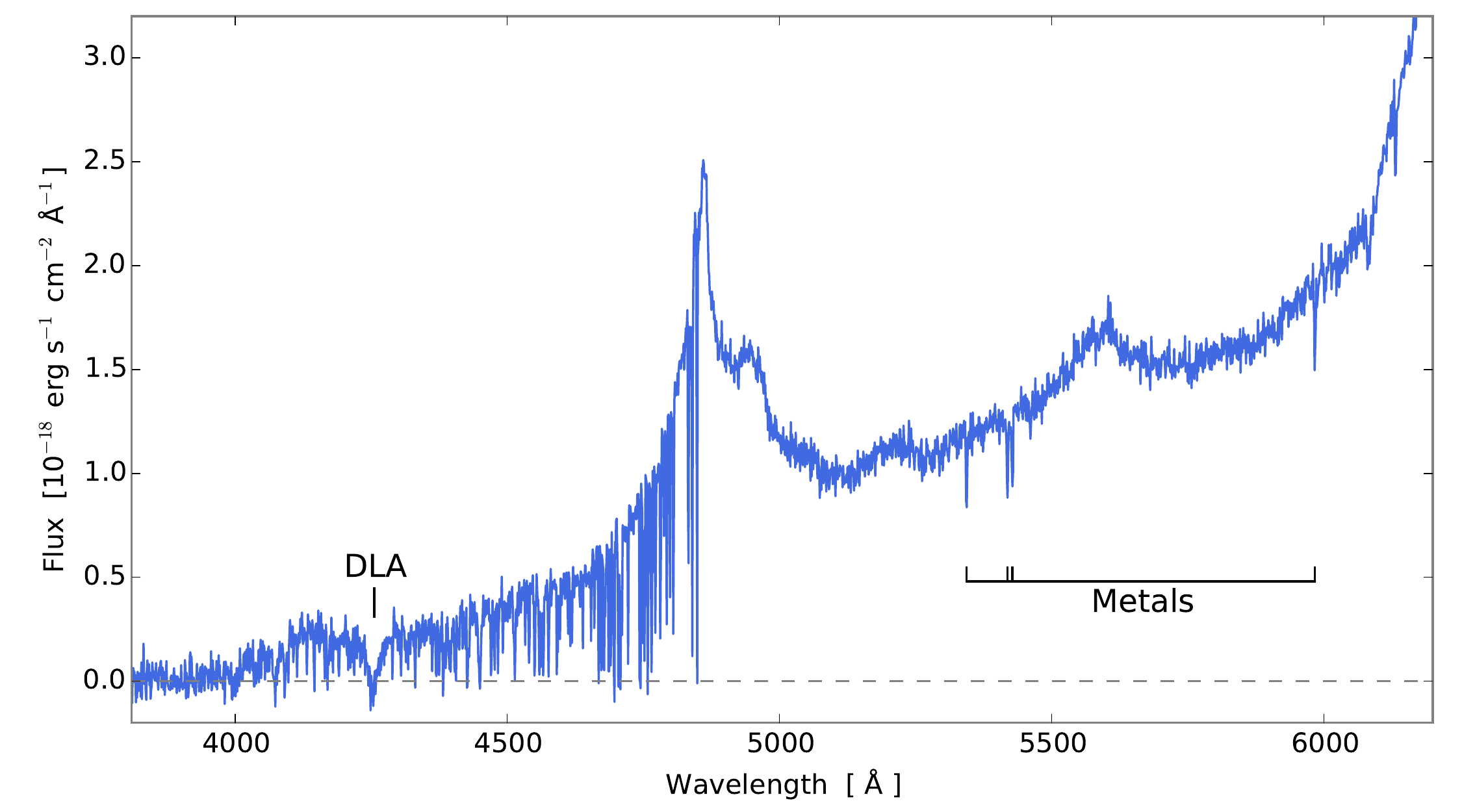}
    \caption{Simulated spectrum of a quasar at $z\approx3$ with a DLA at $z=2.5$. The simulation
    was performed for an intrinsic quasar brightness of $r=19$~mag,
    $A(V)_{\rm DLA}=1.0$ assuming the SMC law, a spectral resolution of $\mathcal{R}=2000$,
    and average signal-to-noise of about 20. The strong \lya\ line as well as associated
    metals lines are clearly visible.
    \label{fig:DLA_simulation}}
\end{figure}

\newpage
Moreover, we have simulated spectra for such reddened quasars with foreground DLAs. One such spectrum is shown in Figure~\ref{fig:DLA_simulation}. It is clear that the high amount of reddening suppresses the blue end of the spectrum significantly and the direct search for the \lya\ absorption line is not feasible for noisy spectra (the spectrum in Figure~\ref{fig:DLA_simulation} has an average signal-to-noise ratio of about 20). Instead, the rest-frame UV metal absorption lines serve as probes for the foreground absorption.
A quasar with an intrinsic brightness of $r=19$, shown in Figure~\ref{fig:DLA_simulation}, has an observed $r$-band magnitude of roughly 23~mag due to the dust reddening from the DLA. Such targets thus require long integration times and/or large telescopes.
This will be the aim of the eHAQ survey, which was initiated earlier this year in March.

The issue of a dust bias in optically selected samples at lower redshifts ($z<1.5$) will also be studied in the future MeerKAT Absorption Line Survey (MALS), which is a large radio survey of the southern hemisphere carried out with the MeerKAT in South Africa. This survey will target 1000 radio and mid-infrared selected quasars, which will have complementary optical spectra. As part of my PhD I have been involved in the optical follow-up for this survey. The goals of MALS will complement the higher redshift eHAQ sample perfectly.

\newpage
In Chapters~\ref{redQSOs} and ~\ref{HAQ}, we identified a large number of missing quasars and a fraction of these had extinction curves steeper than the SMC extinction curve. These were analysed in a separate article, where we have found that the extinction curves towards these quasars seem to be very similar \citep{Zafar2015}. This opens up new questions about the origin of such dust properties: How frequent is the steep extinction curve? Is the excess reddening caused by different grain properties related to the accreting black hole? Or is this due to different chemistry?
The eHAQ survey will provide a larger sample of quasars which we intend to follow-up with the X-shooter spectrograph to study the quasar emission lines in detail.\\

Lastly, in Chapter~\ref{K14}, we presented a spectroscopic sample of quiescent galaxies at $z\sim2$. With the improved redshift information, we could constrain the mass--size relation better, allowing us to study the intrinsic scatter in this relation. We used the scatter as an extra observable to constrain our `dilution'-driven size-evolution model in which increasingly larger galaxies are formed at later times, subsequently quenched and added to the quiescent population. We concluded that although the evolution of the average size of the population was well-explained by our model, the resulting increase in scatter was inconsistent with the data. Therefore, an interplay of various processes (e.g., `dilution', mergers and star formation) is most likely needed to explain the evolution at all times.

%% file: divisions/summary/summary_dk.tex
\chapter*{Resum{\'e} p{\aa} Dansk}

I starten af det 20. {\aa}rhundrede begyndte man at indse, at vort univers ikke blot udg{\o}res af vor egen galakse -- M{\ae}lkevejen -- men i stedet indeholder millioner af galakser ligesom vor egen. Efterh{\aa}nden som vi observerer mere og mere fjerne galakser, b{\aa}de i tid og rum, dukker et {\aa}benlyst sp{\o}rgsm{\aa}l frem: ``Hvordan udviklede de f{\o}rste galakser sig til de magel{\o}se og smukt sammensatte galakser vi ser omringe os i dag?''
Dette er udgangspunktet for denne afhandling.

\subsubsection{D{\ae}mpede Ly\boldmath{$\alpha$} Absorptionslinjesystemer og deres Emissionsmodstykker}
I kapitlerne~\ref{K12} og \ref{K13} analyseres de s{\aa}kaldte ``d{\ae}mpede Lyman-$\alpha$ absorptionslinjesystemer'' (DLA'er), en bestemt type absorptionslinjesystem, med henblik p{\aa} at identificere de galakser, der for{\aa}rsager absorptionen i lyset fra bagvedliggende kvasarer. Ved at lave en model for de galakser, som danner DLA'er, og sammenholde denne med data, b{\aa}de nye og fra litteraturen, bliver f{\o}lgende konklusion draget:

\begin{itemize}
    \item {Fordelingen af st{\o}dparametre og metallicitet for DLA'er er konsistent med forventningen fra den udviklede model, hvori DLA'er er associeret med lyssvage, men ellers almindelige, stjernedannende galakser (Lyman-break galakser).}
\end{itemize}

En af disse DLA'er analyseres i st{\o}rre detalje med data fra ``Very Large Telescope'' (VLT) i Chile og ``{\it Hubble Space Telescope}'' ({\it HST}). I kapitel~\ref{K13} pr{\ae}senteres de nye data, og p{\aa} baggrund af disse konkluderes det at DLA-galaksen er en ung ($\sim100$~Myr), stjernedannende galakse med kraftige udstr{\o}mninger af gas ($v_{\rm ud}\approx160$~\kms). Med de nye data er det muligt for f{\o}rste gang at bestemme massen af stjernerne i en DLA-galakse: $M_{\star} = 2\times10^{9}~M_{\odot}$.

\subsubsection{S{\o}gning efter R{\o}de Kvasarer}
P{\aa} baggrund af observationer, der indikerer, at visse DLA'er indeholder betydelige m{\ae}ngder af st{\o}v, igans{\ae}ttes et st{\o}rre opf{\o}lgende studium af dette st{\o}vs indvirkning p{\aa} vores udv{\ae}lgelsesmetoder af DLA'er. St{\o}vet i DLA'er medf{\o}rer nemlig, at den bagvedliggende kvasars lys bliver svagere. Ydermere {\ae}ndrer st{\o}vet kvasarens spektrale karakteristika, da de korte b{\o}lgel{\ae}ngder undertrykkes mere end de lange. P{\aa} den m{\aa}de kan st{\o}v i DLA'er medvirke til, at kvasarer bagved DLA'er med store m{\ae}ngder st{\o}v (og dermed ogs{\aa} metaller) ikke opdages med traditionelle metoder. I kapitel~\ref{redQSOs} pr{\ae}senteres vore nye udv{\ae}lgelseskriterier, baseret p{\aa} optisk og n{\ae}r-infrar{\o}d fotometri, samt resultaterne af de opf{\o}lgende observationer af 58 kandidater. Ud af disse er 46 (79\%) bekr{\ae}ftet som v{\ae}rende kvasarer. Kun omkring halvdelen af disse 46 r{\o}de kvasarer er identificeret som s{\aa}dan p{\aa} baggrund af den traditionelle metode, som anvendes i ``The Sloan Digital Sky Survey'' (SDSS). Yderligere er kun fire observeret med spektroskopi i SDSS databasen. M{\ae}ngden af st{\o}v, givet som \Av, bestemmes udfra kvasarspektrene, og f{\o}lgende slutninger drages:

\begin{itemize}
    \item {St{\o}vet er prim{\ae}rt for{\aa}rsaget af kvasarens v{\ae}rtsgalakse, og st{\o}vet f{\o}lger i de fleste tilf{\ae}lde kendte r{\o}dfarvningslove udledt for den Lille Magellanske Sky.}
    \item {Der observeres ingen foranliggende absorptionslinjesystemer.}
    \item {For visse kvasarer kan r{\o}dfarvningen ikke beskrives vha. kendte love. I stedet m{\aa} en r{\o}dfarvningslov med stejlere h{\ae}ldning anvendes.}
\end{itemize}

I et fors{\o}g p{\aa} at mindske kontamination fra stjerner og galakser i udv{\ae}lgelsen af kvasarer revideres kriterierne fra kapitel~\ref{redQSOs} efterf{\o}lgende. De reviderede kriterier og observationerne af 159 dermed udvalgte kandidater pr{\ae}senteres i kapitel~\ref{HAQ}: ``{\it The High A(V) Quasar Survey}''. Ud af de 159 kandidater bekr{\ae}ftes 154 (97\%) som v{\ae}rende kvasarer.
I mods{\ae}tning til unders{\o}gelsen i kapitel~\ref{redQSOs} identificeres 30 absorptionslinjesystemer. Det unders{\o}ges derfor om r{\o}dfarvningen af kvasarerne skyldes st{\o}v i kvasaren selv eller i et foranliggende absorptionslinjesystem. Dette testes vha. en statistisk sammenligning af modeller med og uden hypotetiske absorptionslinjesystemer ved r{\o}dforskydning, $z_{\rm abs}$, som giver anledning til r{\o}dfarvning. I afhandlingen benyttes st{\o}rrelsen $A(V)_{\rm abs}$ til at beskrive ekstinktionen i det visuelle $V$-b{\aa}nd for{\aa}rsaget af det foranliggende absorptionslinjesystem. Den anvendte model er beskrevet i detalje i Appendix~\ref{appendix:dust_model}. Ud fra analysen konkluderes f{\o}lgende:

\begin{itemize}
    \item {Ni spektre beskrives bedst med en model hvor st{\o}vet er i et foranliggende absorptionslinjesystem. To af disse bekr{\ae}ftes yderligere af en detektion af et egentligt absorptionslinjesystem ved samme r{\o}dforskydning som modellen forudsiger.}
    \item {Elleve kvasarer har et signifikant overskud af emission ved 2~$\mu$m i kvasarens referenceramme. Dette forklares ved et overskud af emission fra varmt ($T\sim1500$~K) st{\o}v.}
    \item {Sektsen kvasarer har et signifikant underskud af emission ved 2~$\mu$m i kvasarens referenceramme. Ligesom i kapitel~\ref{redQSOs} tyder dette p{\aa} at r{\o}dfarvningsloven i disse tilf{\ae}lde har stejlere h{\ae}ldning end den antagne lov for den Lille Magellanske Sky.}
\end{itemize}

\subsubsection{St{\o}v i D{\ae}mpede Ly\boldmath{$\alpha$} Absorptionslinjesystemer}
En af kvasarerne i HAQ unders{\o}gelsen, HAQ\,2225+0527, havde et foranliggende DLA, men den pr{\ae}limin{\ae}re analyse viste, at r{\o}dfarvningen skyldes st{\o}v i kvasaren selv. I kapitel~\ref{K15} pr{\ae}senteres nye data fra X-shooter-instrumentet p{\aa} VLT. Med de nye data er det muligt at foretage en detaljeret unders{\o}gelse af b{\aa}de DLA og kvasar. De nye analyser viser, at det er st{\o}v i DLA'et, som giver anledning til r{\o}dfarvningen af kvasaren. Ekstinktionen estimeres ud fra spektral modellering: $A(V)_{\rm DLA}=0.28\pm0.01|_{\rm stat}\ \pm0.07|_{\rm sys}$~mag. Dette er konsistent med $A(V)$ bestemt ud fra jern-zink forholdet: $A(V)_{\rm DLA}=0.39^{+0.13}_{-0.10}$~mag.
Den bagvedliggende kvasar er pga. den kraftige r{\o}dfarvning ikke blevet identificeret som en kvasar in SDSS, og dette metal-rige DLA er derfor ikke repr{\ae}senteret i nuv{\ae}rende kataloger fra SDSS. Dette system giver s{\aa}ledes direkte evidens for, at metal-rige DLA'er underrepr{\ae}senteres i optisk udvalgte kataloger.
Derudover observeres s{\aa}kaldt ``delvis d{\ae}kning'' i absorptionslinjerne fra neutralt kulstof og dettes finstrukturlinjer (\ion{C}{i}, \ion{C}{i}$^*$ og \ion{C}{i}$^{**}$). Dette forklares ved, at det absorberende medium har en lille projiceret udstr{\ae}kning i forhold til den bagvedliggende emissionskilde (i dette tilf{\ae}lde kvasarens bredlinje-emissionsregion, BLR).
St{\o}rrelsen af b{\aa}de det absorberende medium og BLR bestemmes til $\sim0.1$~parsec.

\subsubsection{St{\o}rrelsesudviklingen af Kompakte, Tunge Galakser siden R{\o}dforskydning 2}
I det sidste kapitel (kap.~\ref{K14}) studeres de tungeste galakser i det tidlige univers. De f{\o}rste observationer af disse fjerne og r{\o}de galakser (``distant red galaxies'') viste, at disse galakser typisk er 2--6 gange mindre en tilsvarende galakser i det lokale univers ved r{\o}dforskydning 0. Siden da har mange studier forklaret dette som konsekvensen af en r{\ae}kke sammenst{\o}d med mindre galakser.
I kapitel~\ref{K14} pr{\ae}senteres et spektroskopisk udvalg af galakser observeret med Rumteleskopet {\it HST}. Den {\o}gede pr{\ae}cision i bestemmelserne af r{\o}dforskydning g{\o}r det muligt for f{\o}rste gang at unders{\o}ge spredningen i sammenh{\ae}ngen mellem stellar masse og fysisk st{\o}rrelse. Det vises, at st{\o}rrelsesudviklingen kan beskrives som en udvikling af hele ensemblets gennemsnitlige st{\o}rrelse, i stedet for en egentlig udvikling af hver enkel galakse. Denne udvikling drives af dannelsen af st{\o}rre og st{\o}rre galakser ved lavere r{\o}dforskydninger.
Spredningen i relationen kan derimod ikke beskrives korrekt, da denne if{\o}lge modellen vokser med tiden i mods{\ae}tning til data. Det sluttes derfor, at en kombination af forskellige scenarier m{\aa} virke i f{\ae}lleskab for at forklare udviklingen som funktion af kosmisk tid.

%% file: divisions/summary/summary_es.tex
\chapter*{Resumen en Espa{\~n}ol}

\vspace{-5mm}
En el principio del siglo XX fue descubierto que nuestro universo no estaba limitado a nuestra propia galaxia -- La V{\'i}a L{\'a}ctea -- sino que estaba compuesto por millones de galaxias individuales como la nuestra.
Progresivamente cuando observamos galaxias a{\'u}n m{\'a}s lejanas, tanto en el tiempo como en el espacio, aparece una pregunta fundamental: ``?`C{\'o}mo evolucionaron las primeras galaxias a las magn{\'i}ficas y bellas estructuras que vemos en el universo local?'' {\'e}ste es el punto de partida de esta tesis.

\vspace{-2mm}
\subsubsection{Sistemas Amortiguados Ly\boldmath{$\alpha$} y sus Contrapartes de Emisi{\'o}n}

En los Cap{\'i}tulos~\ref{K12} y \ref{K13} se analizan los ``sistemas amortiguados Ly-$\alpha$'' (DLAs), que son una clase de sistemas de absorci{\'o}n, con la finalidad de identificar las galaxias que causan la absorci{\'o}n de la luz de los cu{\'a}sares de fondo. Haciendo un modelo f{\'i}sico de las galaxias que causan los DLAs, y comparando {\'e}ste con los datos, se concluye lo siguiente:

\vspace{-2mm}
\begin{itemize}
    \item {La distribuci{\'o}n de los par{\'a}metros de impacto y metalicidad de los DLAs es consistente con la expectativa del modelo, en el cual los DLAs est{\'a}n asociados con galaxias de baja luminosidad que forman activamente estrellas (llamadas Galaxias Lyman-break).}
\end{itemize}

Uno de estos DLAs se analiza en m{\'a}s detalle con datos del ``Very Large Telescope'' en Chile y del ``{\it Hubble Space Telescope}''. En el Cap{\'i}tulo~\ref{K13} se presentan los nuevos datos, y a la luz de {\'e}stos se concluye que la galaxia asociada con el DLA es joven ($\sim100$~Myr) y tiene una formaci{\'o}n de estrellas activa con efusi{\'o}n potente de gases (con una velocidad de $160$~\kms). Por primera vez es possible determinar la masa de la galaxia asociada con un DLA altamente desplazado hacia el rojo gracias a los nuevos datos de alta calidad. Se mide una masa de $M_{\star}=2\times10^9~M_{\odot}$.

\subsubsection{La b{\'u}squeda de los Cu{\'a}sares Rojos}
En base a observaciones que indican que algunos DLAs contienen polvo se comenz{\'o} un estudio de seguimiento para investigar c{\'o}mo este polvo afecta a nuestros m{\'e}todos de selecci{\'o}n de DLAs. El polvo en ellos tiene el efecto de extinguir los cu{\'a}sares de fondo. Adem{\'a}s se cambian las caracter{\'i}sticas espectrales de los cu{\'a}sares por el polvo, porque las longitudes de onda cortas se suprimen m{\'a}s que las longitudes de onda largas.
De este modo el polvo en los DLAs podr{\'i}a cambiar las caracter{\'i}sticas de los cu{\'a}sares al grado de que no se descubran los cu{\'a}sares por m{\'e}todos tradicionales. Este efecto es m{\'a}s grave para los DLAs que contienen mucho polvo (por lo tanto, tambi{\'e}n muchos metales). En el Cap{\'i}tulo~\ref{redQSOs} se presentan nuestros criterios de selecci{\'o}n nuevos, que se basan en la fotometr{\'i}a {\'o}ptica y del infrarrojo cercano, y los resultados de las observaciones espectrales de seguimiento. De los 58 candidatos que observamos se confirman 46 (79\%) cu{\'a}sares. Los dem{\'a}s son cuatro galaxias rojas, cuatro estrellas fr{\'i}as de tipo M y cuatro objetos que no se identifican con certeza. S{\'o}lo la mitad de los 46 cu{\'a}sares confirmados han sido identificados como tales usando los m{\'e}todos tradicionales, como los que usa el Sloan Digital Sky Survey (SDSS). Adem{\'a}s, s{\'o}lo cuatro est{\'a}n observados con la espectroscop{\'i}a en la base de datos del SDSS. La cantidad de polvo, aqu{\'i} llamado $A(V)$, se determina de los espectros, y se puede concluir lo siguiente:

\begin{itemize}
    \item {El polvo est{\'a} principalmente asociado a la galaxia del cu{\'a}sar, y el polvo est{\'a} descrito por la relaci{\'o}n de enrojecimiento de la Peque{\~n}a Nube de Magallanes (SMC).}
    \item {No se observa ning{\'u}n sistema de lineas de absorci{\'o}n delante de los cu{\'a}sares.}
    \item {El enrojecimiento de algunos cu{\'a}sares no se describe bien por la relai{\'o}n de la SMC. En este caso, se necesita una relaci{\'o}n con pendiente m{\'a}s pronunciada.}
\end{itemize}

Intentando eliminar la contaminaci{\'o}n de estrellas y galaxias en la selecci{\'o}n de cu{\'a}sares, se revisan los criterios posteriormente. Los criterios revisados y las observaciones de los 159 candidatos se presentan en el Cap{\'i}tulo~\ref{HAQ}: ``{\it The High A(V) Quasar Survey}''. De los 159 candidatos se confirman 154 (97\%). En contraste con el estudio en el Cap{\'i}tulo~\ref{redQSOs} se identifican 30 sistemas de l{\'i}neas de absorci{\'o}n. Por lo tanto, se examina si el enrojecimiento est{\'a} causado por el polvo en los cu{\'a}sares o en los sistemas de absorci{\'o}n delante de ellos. {\'e}sto se prueba mediante la comparaci{\'o}n estad{\'i}stica de varios modelos con y sin polvo en un sistema de absorci{\'o}n hipot{\'e}tico. El corrimiento hacia el rojo del sistema de absorci{\'o}n se denota $z_{\rm abs}$, y el monto de extinci{\'o}n causado por el sistema de absorci{\'o}n se denota $A(V)_{\rm abs}$. El modelo se describe en el Ap{\'e}ndice~\ref{appendix:dust_model}. Tras el an{\'a}lisis se concluye:

\begin{itemize}
    \item {Nueve de los espectros son consistentes con el modelo en el cual un sistema de absorci{\'o}n causa el enrojecimiento. Dos de estos se confirman posteriormente por una detecci{\'o}n de un sistema de absorci{\'o}n con el mismo corrimiento hacia el rojo como la expectativa del modelo.}
    \item {Once cu{\'a}sares tienen un exceso significativo en $2~\mu {\rm m}$ en el marco de referencia del cu{\'a}sar. {\'E}sto se explica por un exceso de emisi{\'o}n de polvo caliente ($T\sim1500$~K).}
    \item {Diecis{\'e}is cu{\'a}sares tienen un d{\'e}ficit significativo en $2~\mu {\rm m}$ en el marco de referencia del cu{\'a}sar. Esto indica que la relaci{\'o}n del enrojecimiento tiene una pendiente m{\'a}s pronunciada.}
\end{itemize}

\subsubsection{Polvo en los Sistemas Amortiguados Ly\boldmath{$\alpha$}}
Uno de los cu{\'a}sares en el estudio HAQ, el objeto HAQ 2225+0527, tuvo un DLA delante, pero el an{\'a}lisis principal mostr{\'o} que el enrojecimiento fue causado por polvo en el cu{\'a}sar. En el Cap{\'i}tulo~\ref{K15} se presentan nuevos datos del instrumento X-shooter en el Very Large Telescope. Con estos datos se hace una investigaci{\'o}n m{\'a}s detallada. El nuevo an{\'a}lisis muestra que en realidad es polvo en el DLA que causa el enrojecimiento del cu{\'a}sar. La extinci{\'o}n se calcula con dos m{\'e}todos independientes: Haciendo un an{\'a}lisis espectral se estima $A(V)_{\rm DLA}=0.28\pm0.01|_{\rm stat}\ \pm0.07|_{\rm sys}$~mag, y en base de la relaci{\'o}n de hierro a zinc se determina $A(V)_{\rm DLA}=0.39^{+0.13}_{-0.10}$~mag. Debido al enrojecimiento de la DLA, el cu{\'a}sar detr{\'a}s del DLA no est{\'a} clasificado como tal en SDSS. As{\'i}, este sistema de absorci{\'o}n presta evidencia directa de que los DLAs que contienen muchos metales y polvo est{\'a}n subrepresentados en los cat{\'a}logos actuales seleccionados {\'o}pticamente.
Adem{\'a}s se observa ``cobertura parcial'' en las l{\'i}neas de absorci{\'o}n de carbono neutral y sus niveles de estructura fina (\ion{C}{i}, \ion{C}{i}$^*$ y \ion{C}{i}$^{**}$). La cobertura parcial indica que el medio absorbente es m{\'a}s peque{\~n}o (en proyecci{\'o}n) comparado con la fuente de emisi{\'o}n de fondo (en este caso, la regi{\'o}n de emisi{\'o}n de l{\'i}neas anchas). El tama{\~n}o del medio absorbente y la regi{\'o}n de emisi{\'o}n de l{\'i}neas anchas se determina que es 0.1~p{\'a}rsec.

\subsubsection{La evoluci{\'o}n del Tama{\~n}o de Galaxias Compactas y Masivas}
En el {\'u}ltimo Cap{\'i}tulo (el Cap.~\ref{K14}) se presenta un estudio de las galaxias m{\'a}s masivas y muy corridas hacia el rojo ($z=2$). Las primeras observaciones de estas galaxias rojas y lejanas (llamadas ``distant red galaxies'') han mostrado que las galaxias son 2--6 veces m{\'a}s peque{\~n}as que las galaxias equivalentes en el universo local (es decir, cero corrimiento hacia el rojo). 
Desde entonces muchos estudios han explicado esta evoluci{\'o}n con una secuencia de fusi{\'o}n gal{\'a}ctica con sat{\'e}lites. En el Cap{\'i}tulo~\ref{K14} se presenta una investigaci{\'o}n espectral de galaxias que tienen un corrimiento al rojo de aproximadamente 2 ($z\approx2$) usando el telescopio espacial ``Hubble''.
Con la mejor precisi{\'o}n en la medici{\'o}n del corrimiento al rojo es posible investigar la dispersi{\'o}n en la relaci{\'o}n entre masa estelar y tama{\~n}o. Se muestra que la evoluci{\'o}n de tama{\~n}o puede ser explicado como una evoluci{\'o}n del promedio de la distribuci{\'o}n de tama{\~n}o y no necesariamente un crecimiento de cada galaxia individualmente. Esta evoluci{\'o}n est{\'a} impulsada por la formaci{\'o}n de galaxias progresivamente m{\'a}s grandes hacia bajo corrimiento al rojo. Sin embargo, la dispersi{\'o}n de la relaci{\'o}n no se explica correctamente, ya que {\'e}sta, seg{\'u}n el modelo, crece con el tiempo contrariamente a los datos. Por consiguiente, se concluye que se necesita una combinaci{\'o}n de varios mecanismos juntos para explicar la evoluci{\'o}n en todas {\'e}pocas de la historia c{\'o}smica.

%% file: divisions/appendix/appendix1.tex
\chapter{Appendix to Chapter \ref{intro}}

\renewcommand\theequation{eq. \thesecapp.\arabic{equation}}

\section{Voigt Profile Fitting}
\label{appendix:Voigt}
In order to measure the column densities of metal lines in damped Lyman $\alpha$ absorbers, I fit the absorption lines with Voigt profiles. For that purpose, I wrote my own fitting routine in Python, {\ttfamily\small VoigtLineFit}. The following sections describe the main outline of the software.

\subsection{Absorption Line Profile}
The absorption line arising from a transition $i$ of element $X$ can be described by the line's optical depth, $\uptau$, which is determined by the column density of the element $X$ along with a set of atomic parameters describing the line strength, $f_i$, the damping constant, $\Gamma_i$, and the resonance wavelength, $\lambda_i$, for the transition, $i$:

\begin{equation}
    \uptau _{i, X}(\lambda) = C_i\ N_X\ a_i \ H[a_i, x(\lambda)]
\end{equation}
\noindent
where $C_i$ and $a_i$ are given by:

\begin{equation}
    C_i \equiv \frac {4\ e^2\ \sqrt{\pi^3}} {m_e\ c}\ \frac{f_i}{\Gamma _i}\ \hspace{1cm} \mathrm{and}\ \hspace{1cm} a_i \equiv \frac {\lambda_i\ \Gamma _i} {4 \pi b}
\end{equation}

\noindent
The line profile is determined by the {\it Voigt--Hjerting function}, $H(a_i, x)$:
\begin{equation}
    H(a_i, x) \equiv \frac{a_i}{\pi} \int_{-\infty}^{+\infty} \frac{e^{-y^2}} {(x-y)^2 + a_i^2}\ {\rm d}y
\end{equation}
\noindent
where $x(\lambda) = (\lambda - \lambda_i)/\lambda_D$ is the rescaled wavelength and $y=v/b$ is the velocity of the absorbing atom in units of the broadening parameter, $b$. $\lambda_D = (b/c)\ \lambda_i$ is the Doppler wavelength.

\newpage
Since the integral in the Voigt--Hjerting function is very laborious to evaluate for every iteration in the fit, an analytical approximation is used instead:
\begin{equation}
    H(a_i, x) \approx h - \frac{a_i}{ x^2 \sqrt{\pi}} \left[ h^2\ (4 x^4 + 7 x^2 + 4 + 1.5 x^{-2}) - 1.5 x^{-2} - 1 \right]
\end{equation}
\noindent
where $h=e^{-x^2}$ \citep[see]{TepperGarcia2006, TepperGarcia2007}.
Given the optical depth for the given transition, the resulting flux is given by:
\begin{equation}
    I(\lambda) = I_0(\lambda)\ e^{-\uptau(\lambda)}
\end{equation}
\noindent where $I_0$ is the incident flux. For DLAs, $I_0$ is the spectrum of the background quasar.

\subsection{Fitting Spectra}
The first input required for the fit is obviously the spectral data and information about the spectral resolution. After the data has been loaded, the absorption lines are defined.
The code generates a small (defined in terms of velocity, typically $\pm500$~\kms) region of the spectral data around every single absorption line to be fitted. If spectral regions overlap, they are merged into one region. Next, the velocity components ($z$, $b$) are defined for each line. The code then normalizes each region, if the input data are not normalized, and unwanted parts of the spectra are interactively masked out. Before fitting, linked and fixed parameters are initialized as well as parameter boundaries. The code then fits all the, $N$, regions simultaneously.

For every fit iteration, the code calculates the intrinsic optical depth, $\uptau$ (described above) for the lines defined in each region on the same wavelength grid as the data. The optical depths for all lines in a region are summed and converted to a total intrinsic line profile, $I$. The intrinsic line profile is subsequently convolved with the instrumental line spread function (LSF; assumed to be Gaussian with a width determined by the spectral resolution, $\mathcal{R}$). The $\chi^2$ is calculated as the sum over all regions:
\begin{equation}
    \chi^2 = \sum_{n=1}^N \sum_{m=1}^M \frac{\left( \mathcal{F}_{n,m} - \mathcal{M}_{n,m} \right)^2}{(\sigma_{n,m})^2 + (\sigma_{\rm norm})^2}~,
\end{equation}
\noindent where $\mathcal{F}_{n, m}$ denotes the $m^{\rm th}$ spectral pixel of the data in the $n^{\rm th}$ region, similarly $\mathcal{M}(=I \ast {\rm LFS})$ denotes the model spectrum of the given region, and $\sigma$ refers to the uncertainty of the spectral data. The uncertainty on the continuum normalization for the $n^{\rm th}$ region, $\sigma_{\rm norm}$, is included.
The $\chi^2$ minimization and parameter attributes (e.g., ties and boundaries) are handled by the Python package {\tt lmfit}\footnote{\fontsize{9.4}{10}\selectfont Written by Matthew Newville. Full documentation: http://cars9.uchicago.edo/software/python/lmfit/}.

\newpage

\section{Fitting Dust towards Quasar Sightlines}
\label{appendix:dust_model}
In Chapters~\ref{redQSOs}, \ref{HAQ}, and \ref{K15}, I estimate the amount of dust along quasar sightlines using template fitting. Below I summarize the full model, which is an expanded formalism of the modelling performed in Chapter~\ref{HAQ}.
The model is parametrized by the following set of parameters, $\vec x$:
$$ \vec x = \{z_{\rm abs},\ A(V)_{\rm QSO},\ A(V)_{\rm abs},\ \Delta\beta,\ f_0,\ Fe_2,\ Fe_3 \}\ ,$$
where $z_{\rm abs}$ is the redshift of the absorber; $A(V)_{\rm QSO}$ and $A(V)_{\rm abs}$ denote the $V$-band extinction in the rest-frame of the QSO and absorber, respectively; $\Delta \beta$ denotes the power-law slope relative to the intrinsic slope of the quasar template; $f_0$ is an arbitrary scaling as we do not know the intrinsic brightness of the quasar prior to reddening; $Fe_2$ and $Fe_3$ denote the strengths of the emission template for ${\rm Fe\, \textsc{ii}}$ and ${\rm Fe\, \textsc{iii}}$, respectively.

\noindent
For each set of parameters, the model template is calculated as:

\begin{equation}
    \begin{split}
        T(\vec x, \lambda) = f_0 & \times
        \left[T_{\rm QSO}\cdot \left( \frac{\lambda}{\lambda_0} \right)^{\Delta\beta} +
        Fe_2\cdot T_{\rm Fe \textsc{ii}} + Fe_3\cdot T_{\rm Fe \textsc{iii}} \right] \\
        & \times \, \exp \left[ -\frac{1}{2.5 \log_{10}(e)}
        \left( \xi_{\rm QSO} (\lambda) \cdot A(V)_{\rm QSO} +
        \xi_{\rm abs} (\lambda)\cdot A(V)_{\rm abs} \right) \right]\ ,
    \end{split}
\end{equation}
\noindent
where $T_{\rm QSO}$ denotes the quasar template from Selsing, Fynbo, Christensen, \& Krogager (2015, in preparation) shifted to the redshift of the quasar, $z_{\rm QSO}$, $T_{\rm Fe \textsc{ii}}$ and $T_{\rm Fe \textsc{iii}}$ denote the iron emission template from \citet*{Vestergaard2001}, see Sect.~\ref{appendix:iron_temp}, $\xi_{\rm QSO}$ and $\xi_{\rm abs}$ denote the wavelength dependent reddening curve in the rest-frame of the QSO and absorber, respectively. The reddening curve, $\xi_{\rm abs}$, is therefore implicitly a function of $z_{\rm abs}$.

The $\chi^2$ (which is related to the likelihood, $\mathcal{L}$) is calculated assuming a Gaussian error distribution, neglecting the correlation between neighbouring pixels.

\begin{equation}
    \chi^2 = -2 \ln (\mathcal{L}) = \sum_i^N \frac{\left( D_i - T(\vec x, \lambda_i) \right)^2}
                                                            {\Sigma_i^2}\ .
\label{eq:likelihood}
\end{equation}

The sum is over all $N$ data points $D_i$. The model $T$ is evaluated at the corresponding wavelengths as the data points\footnote{For photometric data, the template is weighted by the appropriate filter transmission curve to generate a synthetic photometric point. For spectra, the template is interpolated onto the same wavelength grid as the observed spectrum.}. Moreover, the errors are convolved with the template error function, $\sigma_T(\lambda)$, which incorporates the quasar-to-quasar variations.

\newpage
The template error function is given as a relative error ($\sigma_T = \frac{\delta T}{T}$, see Figure~\ref{fig:temp_error}). The effective uncertainty for each data point, $\Sigma_i$, is given by adding the intrinsic error ($\sigma_i$) and the template error in quadrature:

\begin{equation}
    \Sigma_i^2 = \sigma_i^2 + \left(D_i \cdot \sigma_T(\lambda_i) \right)^2 \ .
\end{equation}

\noindent
The $\chi^2$ function is minimized with respect to the model parameters, $\vec x$, in order to find the best-fitting parameters. For the minimization, I use the Levenburg--Marquardt algorithm as implemented in the Python package {\tt lmfit}. Since a negative $A(V)$ is non-physical, the parameters $A(V)_{\rm QSO}$ and $A(V)_{\rm abs}$ are only allowed to vary within the given bounds: $A(V) \geq 0$. Similarly, the absorption redshift is kept within $0<z_{\rm abs}<z_{\rm QSO}$. In practice, though, the upper limit on $z_{\rm abs}$ is set to $0.95\times z_{\rm QSO}$ in order to reject absorbers that are very close to the quasar in redshift space. The implementation of the parameter boundaries is handled by {\tt lmfit}.

\subsection{Iron emission template}
\label{appendix:iron_temp}
\begin{figure}
    \centering
    \includegraphics[width=0.98\textwidth]{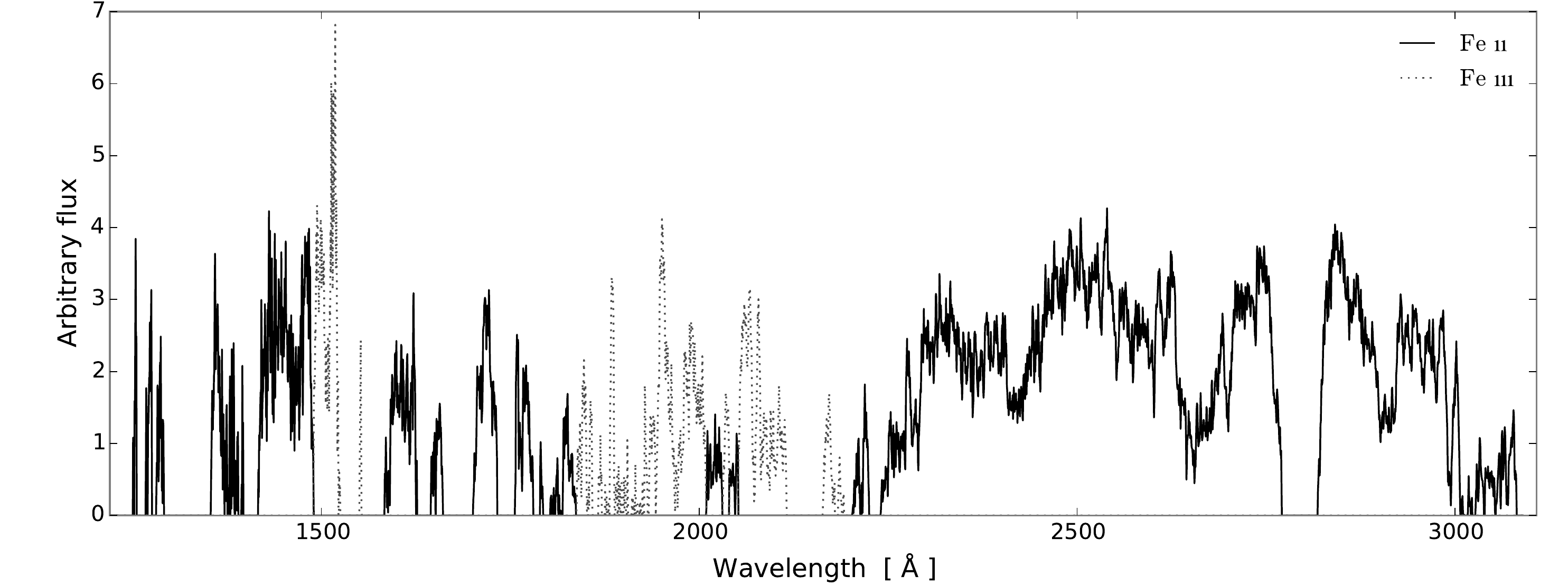}
    \caption{Ultraviolet iron emission template from \citet{Vestergaard2001}.
    \label{fig:iron_temp}}
\end{figure}

The iron emission used together with the quasar template in the model described above is obtained from \citet{Vestergaard2001}. In order to allow the two components (\ion{Fe}{ii} and \ion{Fe}{iii}) to vary independently, the template has been split into two separate templates with varying strengths, $Fe_2$ and $Fe_3$. The two templates are shown in Figure~\ref{fig:iron_temp}. Before introducing the iron template to the model, the template has to be smoothed to match the velocity width of the broad emission lines in the input quasar spectrum.\\
Following Vestergaard \& Wilkes, I convolve the template with a Gaussian kernel of width, ${\rm FWHM_{conv}}$, given by:
$$ {\rm FWHM_{conv}} = \sqrt{\rm (FWHM_{QSO})^2 - (FWHM_{temp})^2}, $$
\noindent
where ${\rm FWHM_{QSO}}$ denotes the full width at half maximum (FWHM) of broad emission lines in the input quasar, and ${\rm FWHM_{temp}}$ refers to the width of the broad emission lines of the quasar used to compile the template ($900$~\kms). The template is rebinned logarithmically to have a constant pixel-size in velocity-space. This keeps the convolution kernel constant across the template. After the convolution, the template is interpolated back onto a linear wavelength grid matching the spectral sampling of the data.

\begin{figure}
    \centering
    \includegraphics[width=0.98\textwidth]{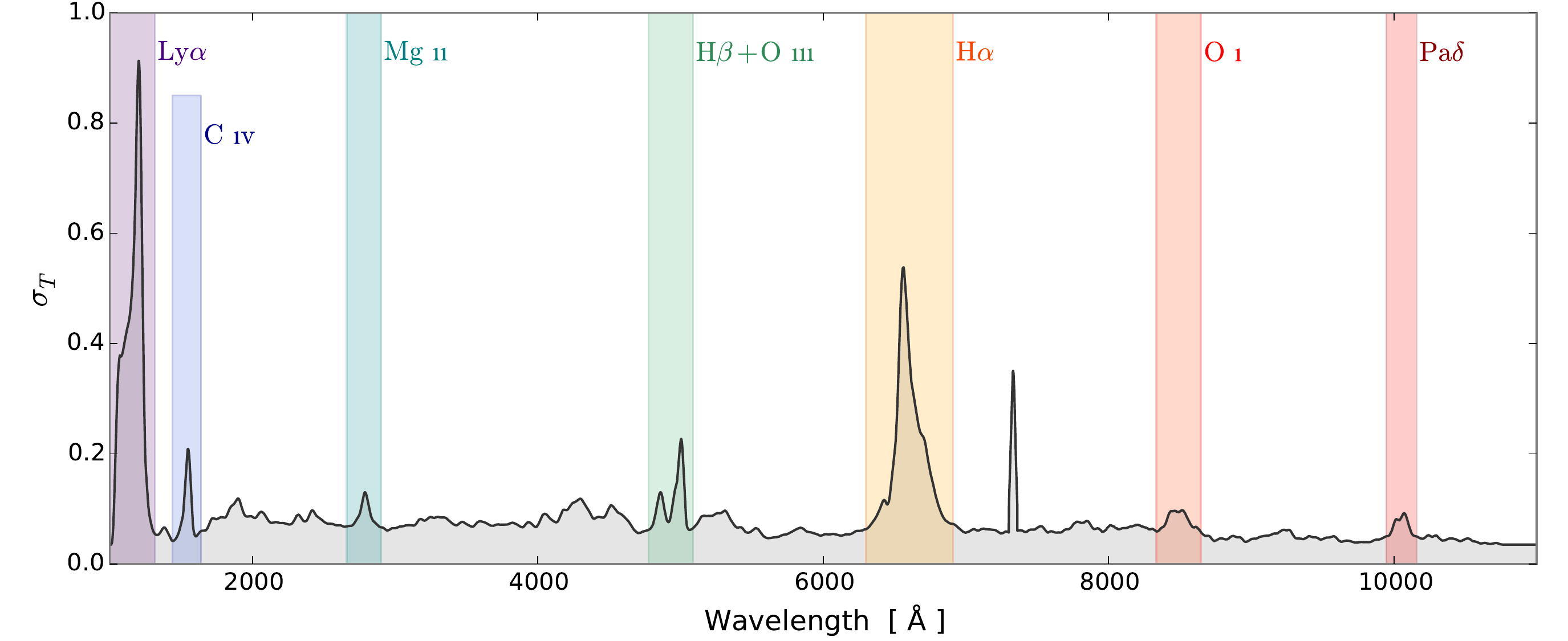}
    \caption{Quasar template error function. The data are provided by
    \citet{Selsing2016}. The most prominent
    emission lines are labeled.
    \label{fig:temp_error}}
\end{figure}

A similar velocity broadening is not necessary for the quasar template, $T_{\rm QSO}$, since the broad emission lines are masked out during the fit. These emission lines are much too variable to provide a good fit with a single template (this is also reflected by the template error function in Figure~\ref{fig:temp_error}).

\subsection{Bayesian Parameter Estimation and Priors}
The inclusion of boundaries in the regular $\chi^2$ minimization leads to incorrect handling of the uncertainties for parameters close to the bounds. Therefore, the fitting code allows a higher degree of accuracy by using a Markov chain Monte Carlo (MCMC) method for the parameter estimation. Also, the MCMC method allows the inclusion of priors on the various parameters. I use the Python package {\tt emcee} \citep{emcee} to sample the posterior probability distribution, $P$:

\begin{equation}
    P(\vec x\, | D_i) = \frac{\mathcal{L}(D_i | \vec x\,)\ p(\vec x)}{p(D_i)}~,
\end{equation}
\noindent
where $\mathcal{L}(D_i | \vec x\,)$ is the likelihood defined in \ref{eq:likelihood}, $p(\vec x)$ refers to the prior probability of the parameters (also simply referred to as the {\it prior} on $\vec x$), and $p(D_i)$ is the probability of the data, $D_i$, which is constant and can be regarded as a normalization constant.

\subsubsection{Setting the priors}
When including a variable slope in the modelling I use a Gaussian prior for the slope: $p(\Delta \beta) = \mathcal{N}(0, 0.2)$. The intrinsic uncertainty on the power-law slope of $0.2$~dex is motivated by previous studies \citep{vandenBerk2001, Krawczyk2015}. For all other parameters, I use na\"ive priors, unless other data suggest differently. For instance, in Chapter~\ref{K15}, I could have included a very restrictive prior on $z_{\rm abs}$, since this quantity is known spectroscopically with very high confidence. I chose, however, to run the chain with this parameter fixed to the spectroscopic value to speed up the process. In the same Chapter, I could include a prior on $A(V)_{\rm abs}$ since this parameter is constrained independently from the abundance ratio analysis. In fact, doing so did not change the best-fitting parameters, since the prior is relatively weak compared to the wealth of data from the full range of the X-shooter spectrum.
Moreover, I include the physically motivated boundaries on $A(V)_{\rm QSO} \geq 0$, $A(V)_{\rm abs} \geq 0$, and $0 < z_{\rm abs} < z_{\rm QSO}$.

\subsubsection{Initiating the chain}
The software {\tt emcee} uses a set of so-called `walkers' to investigate the parameter space. Each walker is basically its own chain, but the behaviour of one walker depends on the position of the other walkers. The walkers need to be initiated at some location in the $N_{\rm par}$ dimensional parameter space (here $N_{\rm par}=7$). From that initial location, the walkers branch out and sample the posterior probability for $n$ iterations. The first phase of the sampling is referred to as the `burn in' phase where the walkers are exploring a large part of parameter space. As the chain progresses, the walkers converge towards the maximum likelihood. Hereafter, the walkers sample the posterior probability and the recovered samples are used to infer the posterior probability distribution for the parameters. The `burn in' samples are rejected from the chain, since they are not representative of the posterior distribution.

\begin{figure}
\begin{center}
    \includegraphics[width=0.90\textwidth]{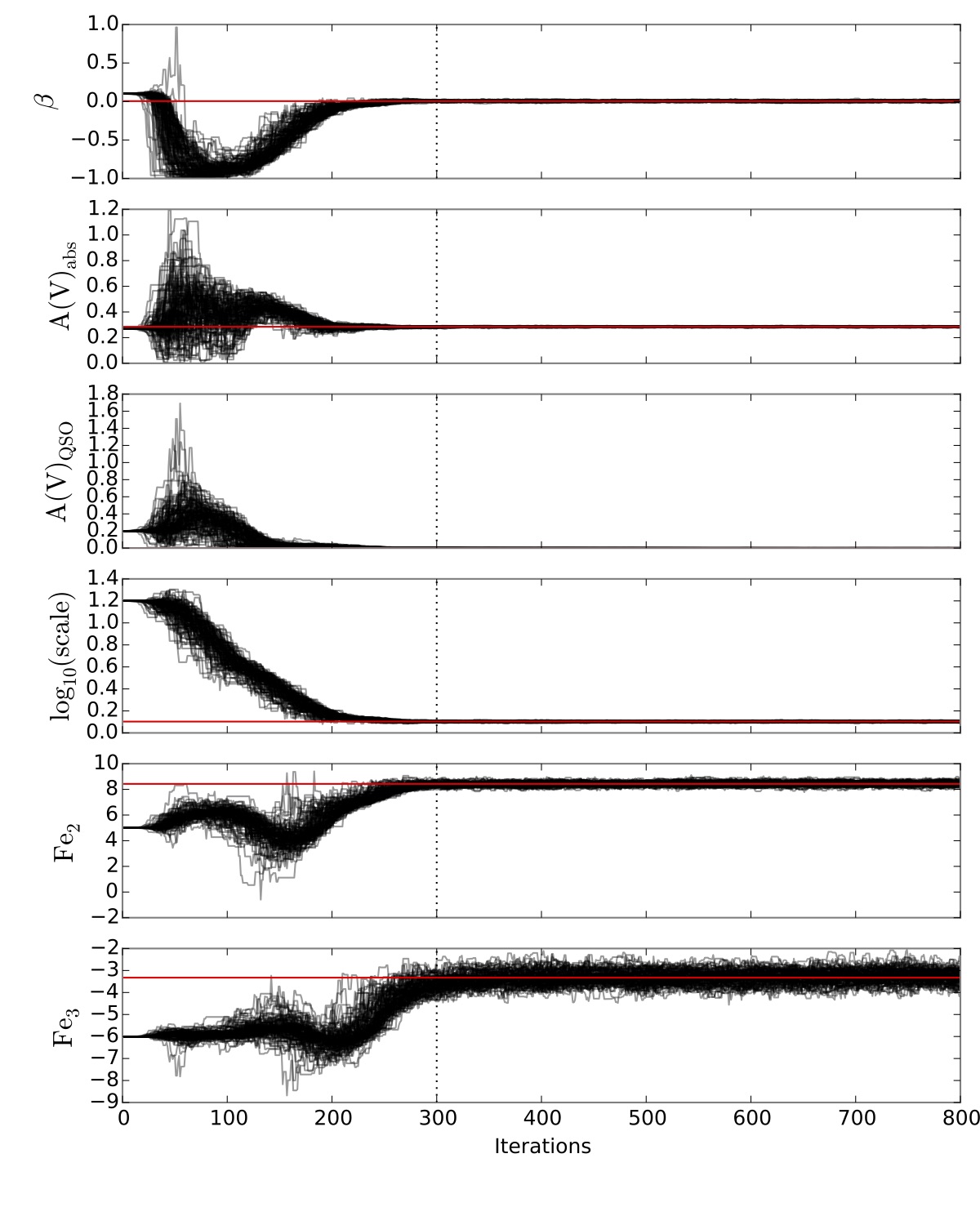}
    \caption{The MCMC chain from {\tt emcee} for each parameter.
    The best-fit parameters are indicated in each panel with red lines.
    The first 300 iterations (dotted vertical line) are discarded as `burn in' phase.}
    \label{fig:MCMC_chain}
\end{center}
\end{figure}

\begin{figure}
\begin{center}
    \includegraphics[width=1.0\textwidth]{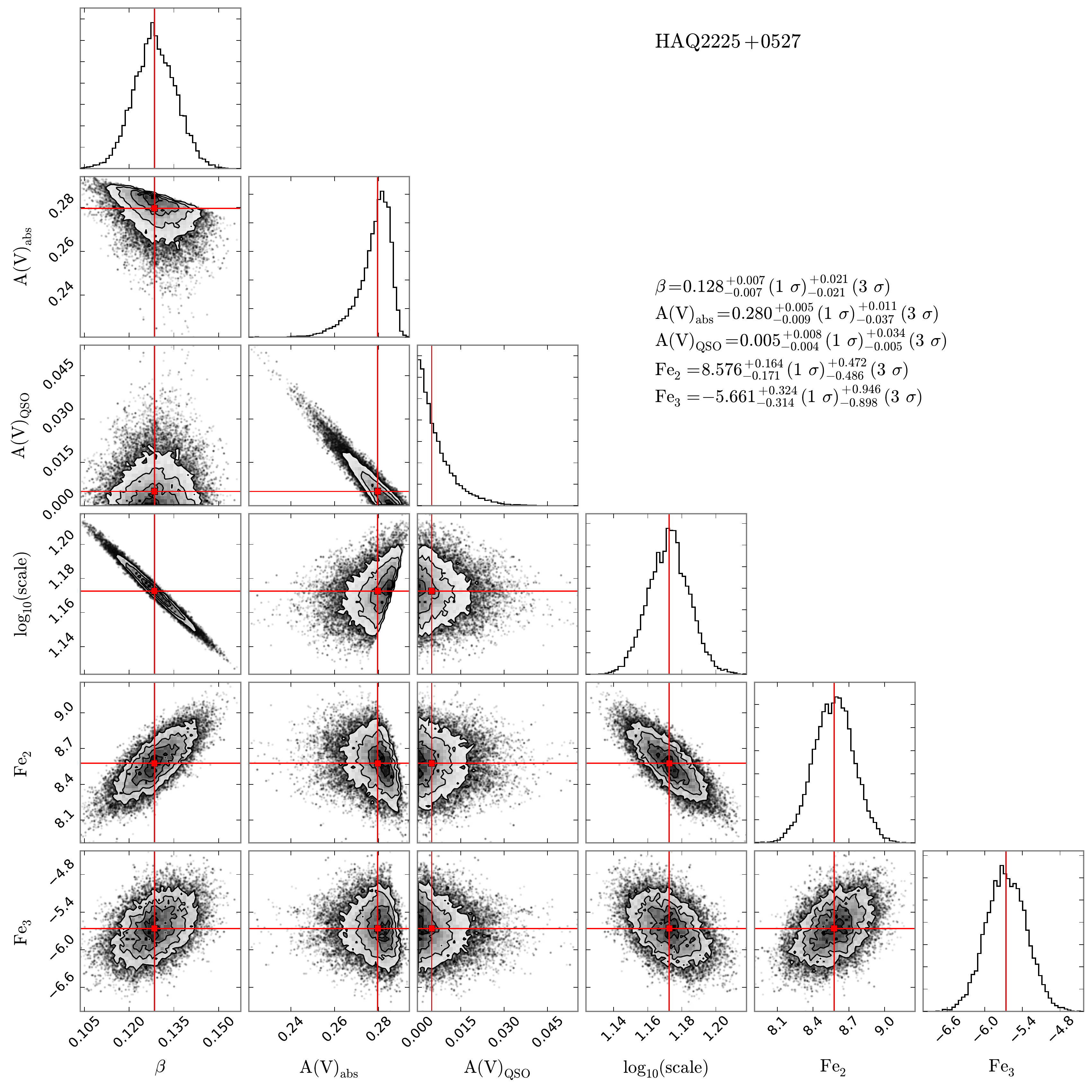}
    \caption{Marginalized posterior probabilities and the correlations
    among various parameters. The best-fit parameters are stated in the
    upper right corner. These are indicated in the figure with red lines.}
    \label{fig:posterior}
\end{center}
\end{figure}

In Chapter~\ref{K15}, I initiate the sampler with 100 walkers and run the chain for 800 iterations, 300 of which are removed as `burn in', see Figure~\ref{fig:MCMC_chain}. The walkers are initiated around the best-fitting solution from the simple $\chi^2$ minimization. The results for the quasar J\,2225+0527 analysed in Chapter~\ref{K15} are shown in Figure~\ref{fig:posterior}. The best-fitting parameters are quoted as the median of the marginalized posterior with 1 (3) sigma uncertainty given by the 16 and 84 (0.2 and 99.8) percentiles.

%% file: divisions/appendix/publications.tex
\chapter{List of Publications}

Number of refereed first-author articles: 5\ ---\ Citations: 80.\\
Number of refereed articles: 16\ ---\ Citations: 428.

\section*{Refereed Publications}
\begin{enumerate}

  \item{ Heintz,~K.~E.; Fynbo,~J.~P.~U.; {\bf Krogager,~J.-K.}; Vestergaard,~M.; Møller,~P.;
  Arabsalmani,~M.; Geier,~S.; Noterdaeme,~P.; Ledoux,~C.; Saturni,~F.~G.; Venemans,~B.,\\
  ``{Serendipitous discovery of a projected pair of QSOs separated by 4.5 arcsec on the sky},''\\
  accepted for publication in {\it Astronomical Journal, 2016. arXiv:1604.01361}
  }
    
  \item{ Selsing,~J.; Fynbo,~J.~P.~U.; Christensen,~L.; \& {\bf Krogager,~J.-K.},\\
  ``{An X-Shooter composite of bright $1<z<2$ quasars from UV to infrared},''\\
  {\it Astronomy \& Astrophysics, 2016, vol. 585, id. 87, 13 pp.}
  }
  
  \item {{\bf Krogager, J.-K.}; Fynbo, J.~P.~U.; Noterdaeme, P.; Zafar, T.; M{\o}ller, P.;
    Ledoux, C.; Kr{\"u}hler, T.;\\ \& Stockton, A.,\\
    ``A quasar reddened by a sub-parsec-sized, metal-rich and dusty cloud in a damped Lyman
    $\alpha$ absorber at $z=2.13$,''\\
    {\it Monthly Notices of the Royal Astronomical Society, 2016, vol. 455, pp. 2698-2711.}
  }

  \item{ Zafar,~T.; M{\o}ller,~P.; Watson,~D.; Fynbo,~J.~P.~U.;
  {\bf Krogager,~J.-K.}; Zafar,~N.; Saturni,~F.~G.; Geier,~S.;
  \& Venemans,~B.~P.,\\
  ``{Extinction curve template for intrinsically reddened quasars},''\\
  {\it Astronomy \& Astrophysics, 2015, vol. 584, id. 100, 8 pp.}
  }

  \item{ {Hartoog},~O.~E.; {Malesani},~D.; {Fynbo},~J.~P.~U.; {Goto},~T.; {Kr{\"u}hler},~T.;
  {Vreeswijk},~P.~M.; {De~Cia},~A.; {Xu},~D.; {M{\o}ller},~P.; {Covino},~S.; {D'Elia},~V.;
  {Flores},~H.; {Goldoni},~P.; {Hjorth},~J.; {Jakobsson},~P.;\\
  {\bf{Krogager},~J.-K.}; {Kaper},~L.; {Ledoux},~C.; {Levan},~A.~J.; {Milvang-Jensen},~B.;
  {Sollerman},~J.; {Sparre},~M.; {Tagliaferri},~G.; {Tanvir},~N.~R.; {de~Ugarte~Postigo},~A.;
  {Vergani},~S.~D.; {Wiersema},~K.; {Datson},~J.; {Salinas},~R.; {Mikkelsen},~K.;
  \& {Aghanim},~N.,\\
  ``{VLT/X-shooter spectroscopy of the afterglow of the Swift GRB 130606A:
  Chemical abundances and reionisation at $z\sim 6$},''
  {\it Astronomy \& Astrophysics, 2015, vol. 580, id. 139, 15 pp.}
  }

\newpage

  \item { {\bf Krogager, J.-K.}; {Geier}, S.; {Fynbo}, J.~P.~U.; {Venemans}, B.~P.;
  {Ledoux}, C.; {M{\o}ller}, P.;\\
  {Noterdaeme}, P.; {Vestergaard}, M.; {Kangas}, T.; {Pursimo}, T.; {Saturni}, F.~G.;
  \& {Smirnova}, O.,\\
  ``{The High A$_{V}$ Quasar Survey: Reddened quasi-stellar objects selected from
  optical/near-infrared photometry -- II},''
  {\it The Astrophysical Journal Supplement Series, 2015, vol. 217, id. 5, 26 pp.}
  }

  \item{ {\bf Krogager, J.-K.}; {Zirm}, A.~W.; {Toft}, S.; {Man}, A.; \&
	{Brammer}, G.,\\
	``{A spectroscopic sample of massive, quiescent $z\sim2$ galaxies:
	implications for the evolution of the mass--size relation},''
	{\it The Astrophysical Journal, 2014, vol. 797, id. 17, 14pp.}
  }

  \item{ {Toft}, S.; {Smol{\v c}i{\'c}}, V.; {Magnelli}, B.; 
	{Karim},~A.; {Zirm},~A.; {Michalowski},~M.; {Capak},~P.; 
	{Sheth},~K.; {Schawinski},~K.; {\bf Krogager,~J.-K.}; {Wuyts},~S.;
	{Sanders},~D.; {Man},~A.~W.~S.; {Lutz},~D.; {Staguhn},~J.;
	{Berta},~S.; {Mccracken},~H.; {Krpan},~J.; \& {Riechers},~D.,\\
	``{Submillimeter galaxies as progenitors of compact quiescent galaxies},''
	{\it The Astrophysical Journal, 2014, vol. 782, id. 68, 12pp.}
  }

  \item{ {Fynbo},~J.~P.~U.; {Geier},~S.~J.; {Christensen},~L.; 
	{Gallazzi},~A.; {\bf Krogager,~J.-K.}; {Kr{\"u}hler},~T.; 
	{Ledoux},~C.; {Maund},~J.~R.; {M{\o}ller},~P.; {Noterdaeme},~P.;
	{Rivera-Thorsen},~T.; \& {Vestergaard}, M.,\\
	``{On the two high-metallicity DLAs at z = 2.412 and 2.583 towards Q 0918+1636},''
	{\it Monthly Notices of the Royal Astronomical Society, 2013, vol. 436, pp. 361--370.}
  }

  \item{ {\bf Krogager,~J.-K.}; {Fynbo},~J.~P.~U.; {Ledoux},~C.;
	{Christensen},~L.; {Gallazzi},~A.; {Laursen},~P.; {M{\o}ller},~P.;
	{Noterdaeme},~P.; {P{\'e}roux},~C.; {Pettini},~M.; \& 
	{Vestergaard}, M.},\\
	``{Comprehensive study of a z = 2.35 DLA galaxy: mass, metallicity, age,
	morphology and SFR from HST and VLT}'',
	{\it Monthly Notices of the Royal Astronomical Society, 2013,
	vol. 433, pp. 3091--3102.
  }

  \item{ {Dahle}, H.; {Gladders},~M.~D.; {Sharon},~K.; {Bayliss},~M.~B.;
	{Wuyts},~E.; {Abramson},~L.~E.; {Koester},~B.~P.; {Groeneboom},~N.; 
	{Brinckmann},~T.~E.; {Kristensen},~M.~T.; {Lindholmer},~M.~O.;
	{Nielsen},~A.;\\
	{\bf Krogager,~J.-K.}; \& {Fynbo}, J.~P.~U.,\\
	``{SDSS J2222+2745: A Gravitationally lensed sextuple quasar with a
	maximum image separation of 15\farcs1 discovered in the
	Sloan Giant Arcs Survey}'',\\
	{\it The Astrophysical Journal, 2013, vol. 773, id. 146, 10pp.} 
  }

  \item{ {Ilbert},~O.; {McCracken},~H.~J.; {Le~F{'e}vre},~O.; {Capak},~P.; {Dunlop},~J.; {Karim},~A.; {Renzini},~M.~A.; {Caputi},~K.; {Boissier},~S.; {Arnouts},~S.; {Aussel},~H.; {Comparat},~J.; {Guo},~Q.; {Hudelot},~P.; {Kartaltepe},~J.; {Kneib},~J.~P.; {\bf{Krogager},~J.~K.}; {Le~Floc'h},~E.; {Lilly},~S.; {Mellier},~Y.; {Milvang-Jensen},~B.; {Moutard},~T.; {Onodera},~M.; {Richard},~J.; {Salvato},~M.; {Sanders},~D.~B.; {Scoville},~N.; {Silverman},~J.~D.; {Taniguchi},~Y.; {Tasca},~L.; {Thomas},~R.; {Toft},~S.; {Tresse},~L.; {Vergani},~D.; {Wolk},~M.; \& {Zirm},~A.,\\
  ``{Mass assembly in quiescent and star-forming galaxies
  since $z\sim4$ from UltraVISTA},''\\
  {\it Astronomy \& Astrophysics, 2013, vol. 556, id. 55, 19 pp.}
  }

  \item{ {Fynbo},~J.~P.~U.; {\bf{Krogager},~J.-K.}; {Venemans},~B.; {Noterdaeme},~P.; {Vestergaard},~M.; {M{\o}ller},~P.; {Ledoux},~C.;
  \& {Geier},~S.,\\
  ``{Optical/near-infrared selection of red quasi-stellar objects: Evidence for steep extinction curves toward galactic centers?},''\\
  {\it The Astrophysical Journal Supplement Series, 2013, vol. 204, id. 6, 14 pp.}
  }
  
  \item{ {\bf{Krogager},~J.-K.}; {Fynbo},~J.~P.~U.; {M{\o}ller},~P.; {Ledoux},~C.; {Noterdaeme},~P.; {Christensen},~L.; {Milvang-Jensen},~B.; \& {Sparre},~M.,\\
  ``{On the sizes of $z{\gtrsim}2$ damped Ly{$\alpha$} absorbing galaxies},''
  {\it Monthly Notices of the Royal Astronomical Society: Letters,
  2012, vol. 424, pp. L1--L5.}
  }

  \item{ {Fynbo},~J.~P.~U.; {Ledoux},~C.; {Noterdaeme},~P.; {Christensen},~L.; {M{\o}ller},~P.; {Durgapal},~A.~K.; {Goldoni},~P.; {Kaper},~L.; {\bf{Krogager},~J.-K.}; {Laursen},~P.; {Maund},~J.~R.; {Milvang-Jensen},~B.; {Okoshi},~K.; {Rasmussen},~P.~K.; {Thorsen},~T.~J.; {Toft},~S.; \& {Zafar},~T.,\\
  ``{Galaxy counterparts of metal-rich damped Ly{$\alpha$} absorbers -- II. A solar-metallicity and dusty DLA at $z_{\rm abs}= 2.58$},''
  {\it Monthly Notices of the Royal Astronomical Society, 2011, vol. 413, pp. 2481--2488.}
  }
  
  \item{{Cupani},~G.; {Cristiani},~S.; {D'Odorico},~V.; {Milvang-Jensen},~B.; \& {\bf{Krogager},~J.-K.},\\
  ``{When two become one: an apparent QSO pair turns out to be a
  single quasar},''\\
  {\it Astronomy \& Astrophysics, 2011, vol. 529, id. 99, 3 pp.}
  }

\end{enumerate}